\documentclass[journal]{IEEEtran}
\usepackage{amsmath,amsfonts,bm}
\usepackage{algorithm}
\usepackage{algpseudocode}
\usepackage{array}
\usepackage[caption=false,font=normalsize,labelfont=sf,textfont=sf]{subfig}
\usepackage{textcomp}
\usepackage{stfloats}
\usepackage{url}
\usepackage{verbatim}
\usepackage{graphicx}
\usepackage{cite}
\usepackage{placeins}
\usepackage{acronym}
\acrodef{elbo}[ELBO]{evidence lower bound}
\acrodef{sbl}[SBL]{sparse Bayesian learning}
\acrodef{map}[MAP]{maximum a-posteriori}
\acrodef{ml}[ML]{maximum likelihood}
\acrodef{kl}[KL]{Kullback-Leibler}
\acrodef{pdf}[PDF]{probability density function}
\acrodef{lse}[LSE]{line spectral estimation}
\acrodef{snr}[SNR]{signal-to-noise ratio}
\acrodef{sir}[SIR]{signal-to-interference ratio}
\acrodef{em}[EM]{expectation-maximization}
\acrodef{awgn}[AWGN]{additive white Gaussian noise}
\acrodef{ospa}[OSPA]{optimal subpattern assignment}
\acrodef{gospa}[GOSPA]{generalized optimal subpattern assignment}
\acrodef{mca}[MCA]{morphological component analysis}
\acrodef{aaf}[AAF]{anti-aliasing filter}
\acrodef{if}[IF]{intermediate frequency}
\acrodef{fmcw}[FMCW]{frequency-modulated continuous wave}
\acrodef{crlb}[CRLB]{Cramer-Rao lower bound}
\acrodef{mse}[MSE]{mean squared error}
\acrodef{rmse}[RMSE]{root mean squared error}
\acrodef{dft}[DFT]{discrete Fourier transform}
\acrodef{rd}[RD]{range-Doppler}
\acrodef{pmcw}[PMCW]{phase-modulated continuous wave}
\acrodef{ofdm}[OFDM]{orthogonal frequency division multiplexing}

\usepackage{pgfplots}
\pgfplotsset{compat=newest}
\usepgfplotslibrary{groupplots}
\usepgfplotslibrary{statistics}
\usetikzlibrary{spy}
\usetikzlibrary{shapes,patterns,svg.path,arrows.meta,calc,positioning,decorations.pathmorphing}
\pgfdeclarelayer{background}
\pgfdeclarelayer{foreground}
\pgfsetlayers{background,main,foreground}
\tikzset{snake it/.style={decorate, decoration=snake}}
\tikzset{>=stealth}

\usepackage{makecell}

\usepackage[hidelinks]{hyperref}

\usepackage{color}
\makeatletter
\def\mathcolor#1#{\@mathcolor{#1}}
\def\@mathcolor#1#2#3{%
  \protect\leavevmode
  \begingroup
    \color#1{#2}#3%
  \endgroup
}
\makeatother

\newcommand{\expect}[2]{{\bigl\langle{#1}\bigr\rangle}_{#2}}
\DeclareMathOperator*{\argmax}{arg\,max}
\DeclareMathOperator*{\argmin}{arg\,min}
\renewcommand{\arraystretch}{1.5}

\providecommand{\ist}{\hspace*{.3mm}}

\providecommand{\rmv}{\hspace*{-.3mm}}
\providecommand{\iist}{\hspace*{1mm}}

\providecommand{\nn}{\nonumber}

\begin{document}

\title{Variational Signal Separation for Automotive Radar Interference Mitigation}

\author{Mate Toth, Erik Leitinger, and Klaus Witrisal
\thanks{M.\ Toth and K.\ Witrisal are with the Institute of Communication Networks and Satellite Communications, Graz University of Technology, Graz, Austria (e-mail: (mate.a.toth,witrisal)@tugraz.at). E.\ Leitinger is with the Laboratory of Signal Processing and Speech Communication, Graz University of Technology, Graz, Austria (e-mail: erik.leitinger@tugraz.at). E.\ Leitinger and K.\ Witrisal are further associated with the Christian Doppler Laboratory for Location-aware Electronic Systems. This work was supported in part by the Christian Doppler Research Association.}
\vspace*{-5mm}}

\maketitle
\begin{abstract}
Algorithms for mutual interference mitigation and object parameter estimation are a key enabler for automotive applications of \ac{fmcw} radar. In this paper, we introduce a signal separation method to detect and estimate radar object parameters while jointly estimating and successively canceling the interference signal. The underlying signal model poses a challenge, since both the coherent radar echo and the non-coherent interference influenced by individual multipath propagation channels must be considered. Under certain assumptions, the model is described as a superposition of multipath channels weighted by parametric interference chirp envelopes. Inspired by \ac{sbl}, we employ an augmented probabilistic model that uses a hierarchical Gamma-Gaussian prior model for each multipath channel. Based on this, an iterative inference algorithm is derived using the variational \ac{em} methodology. The algorithm is statistically evaluated in terms of object parameter estimation accuracy and robustness, indicating that it is fundamentally capable of achieving the \ac{crlb} with respect to the accuracy of object estimates and it closely follows the radar performance achieved when no interference is present.
\end{abstract}
\acresetall

\begin{IEEEkeywords}
	line spectral estimation, sparse Bayesian learning, variational \ac{em}, signal separation, automotive radar, interference.
\end{IEEEkeywords}

\section{Introduction}
\label{sec:introduction}

The growing number of radar-equipped vehicles is expected to cause mutual interference, potentially compromising safety-critical automotive radar applications~\cite{kunert_report_2010, roos_radar_2019}. As a result, this issue has received significant research attention in the past decade. Large industry research projects have been conducted to analyze interference mitigation methods and strategies for using the automotive radar spectrum~\cite{kunert_eu_2012, consortium_gesamtbewertung_2022}. From an academic perspective, the interference problem presents a challenge in signal modeling and processing, making it an interesting area for algorithm design.  Early analyses were conducted in~\cite{hischke_collision_1995, tullsson_topics_1997, brooker_mutual_2007}, and since then, a significant amount of research has been published on the subject. 

\subsection{State of the Art}
\label{subsec:SOTA}

Interference modeling efforts can be categorized into scenario-level and signal-level models. Works in the former category characterize the probability of interference occurring and the received interference power, whereas the latter models the specific form of interference in the received radar signal. Scenario-level modeling approaches include analyses via stochastic geometry~\cite{hourani_stochastic_2017}, graph theory~\cite{torres_analysis_2022} as well as other extensive numerical simulation frameworks~\cite{schipper_simulative_2015}. Signal-level models have been developed for different modulation schemes, including the most commonly used \ac{fmcw}~\cite{toth_analytical_2018,kim_peerpeer_2018} as well as \ac{pmcw}~\cite{bourdoux_pmcw_2021} and \ac{ofdm}~\cite{schweizer_mutual_2021} radars. Experimental evaluations can be found e.g. in~\cite{rameez_experimental_2020, ossowska_imiko_2020}. Recent modeling efforts also consider the interference propagation channel~\cite{torres_channel_2020, cardona_integrating_2023}. This paper focuses on the mutual interference of \ac{fmcw} radars, taking into account the interference channel.

Approaches to mitigate interference include avoidance strategies using cognitive radar principles~\cite{hakobyan_interferenceaware_2019, tovar_torres_automotive_2023} as well as exploiting vehicle-to-vehicle communication~\cite{khoury_radarmac_2016, aydogdu_radchat_2021}. In terms of signal processing algorithms to cancel the interference in the receiver, a vast number of algorithms have been proposed. As mutual \ac{fmcw} interference is commonly of a short duration as compared to the total number of received samples, \textit{zeroing}, optionally followed by interpolating the zeroed samples, has been discussed first in~\cite{tullsson_topics_1997}. This method typically necessitates a separate prior interference detection step, e.g.~\cite{fischer_robust_2015, liu_peltkcn_2020, pernstal_gip_2020, shimura_advanced_2022}. Interpolation approaches include autoregressive modeling~\cite{rameez_autoregressive_2020}, time-frequency interpolation~\cite{neemat_interference_2019} and sparsity-based algorithms~\cite{bechter_automotive_2017, chen_automotive_2020, wang_matrixpencil_2021}. Alternative approaches to interpolation include spatial beamforming~\cite{bechter_blind_2018, rameez_adaptive_2018}, adaptive noise cancellers~\cite{jin_automotive_2019, wang_dual_2023}, time-frequency~\cite{rameez_interference_2022, toth_slowtime_2020} and nonlinear filtering~\cite{wagner_thresholdfree_2018, muja_interference_2022}. Denoising based on signal decomposition and thresholding has been proposed in e.g.~\cite{wu_iterative_2020}. Additionally, many works are making use of deep learning~\cite{oyedare_interference_2022} to detect~\cite{hille_fmcw_2022} and mitigate interference~\cite{rock_resourceefficient_2021,fuchs_automotive_2020,oswald_end--end_2023}.

If the interference signal is estimated, it may be canceled completely by subtraction. Proposed methods for this include~\cite{bechter_estimation_2017, correas_sparse_2019, torres_automotive_2023}. However, if the useful radar return itself, here termed the \textit{object signal}, is also explicitly considered in the estimation, the problem can be cast as \textit{signal separation}. This problem is generally ill-posed, but can be solved by assuming that the object and interference signal components are represented by a \textit{sparse} basis in different signal bases. In~\cite{uysal_synchronous_2018}, the author applied the framework of \ac{mca}~\cite{starck_redundant_2004} modeling the interference and object signals as being sparse in the discrete Fourier and discrete short-time Fourier domains, respectively. This assumption allows for the joint sparse recovery of the signals to be written as a modified dual \textit{basis pursuit} optimization problem~\cite{chen2001SIAMRev}, for which an iterative algorithm was proposed.
Similar algorithms can be found in~\cite{xu_interference_2021, xu_bilevel_2022, wang_interference_2021, wang_mutual_2023}.

\subsection{Contributions and Notations}
\label{subsec:contribution}

Although the existing literature is extensive, we identify the following open topics for which our work provides novel insights. Previously proposed algorithms are applied as \textit{preprocessing} in a conventional \textit{range-Doppler processing} chain. This approach is practical, but has inherent limitations. In general, the goal of radar sensing is to provide estimates of object parameters, which is equivalent to estimating the parameters of a multipath propagation channel containing static (such as buildings, traffics signs or trees) and moving objects (such as vehicles, bicycle or pedestrians). Note that parametric channel models typically represent multipath propagation as a linear superposition of weighted Dirac delta distributions - or spectral lines - with distinct supports in the underlying dispersion domain (range, angle of arrival, angle of departure, Doppler frequency, and combinations thereof). Therefore, estimating multipath parameters can be cast as a \ac{lse} problem~\cite{grebien_super-resolution_2024}, for which \ac{sbl}~\cite{tipping1999NeurIPS:RelevanceVector,tipping2003WAIS:FastMarginalSparseBayesian} and related algorithms have been developed. An \ac{sbl}-type method has been previously proposed as a preprocessing step to interpolate the received signal after interference zeroing~\cite{chen_automotive_2020}, as referenced in Section~\ref{subsec:SOTA}. Our proposed algorithm's design and its aims are however distinct, as detailed in the remainder of this paper.

To achieve robust inference performance, it is necessary to consider the entire problem of parameter estimation under the influence of interfering signals. Previous methods are also often based on heuristics or require the prior setting of certain algorithm parameters. This makes it necessary to manually fine-tune the range-Doppler processing chain and makes it difficult to give performance guarantees. Our proposed algorithm is derived within a rigorous theoretical framework. Specific contributions are the following:
\vspace*{-1mm}
\begin{itemize}
\item We develop a new probabilistic signal model for the problem of mutual \ac{fmcw} radar interference. The model incorporates the multipath propagation as line spectra for both the coherently received radar echo and the interference.
\item We propose a novel inference algorithm inspired by \ac{sbl}~\cite{tipping2003WAIS:FastMarginalSparseBayesian ,shutin2011TSP:fastVSBL, shutin2013:VSBL} for \ac{lse} that is able to infer the superposition of sparse line spectra. The algorithm is based on the variational \ac{em} approach. It jointly estimates the objects' range-Doppler parameters with the parameters of the interference signal and multipath channel. This makes it explicitly robust to mutual interference. Note that the object parameters are estimated super-resolved in a grid-less manner~\cite{grebien_super-resolution_2024}.
\item We quantitatively show near-optimal multi-target detection and parameter estimation performance by statistically evaluating the proposed method in simulation, comparing it to the \ac{crlb} as well as to the interference-free case. We also showcase that the algorithm significantly outperforms a few established signal preprocessing methods for interference mitigation.
\item We include investigations on difficult-to-handle scenarios of model mismatch and poor signal separability. We point out the relative resilience of the algorithm to these challenges, and provide an initial proof of concept based on measurement data.
\end{itemize}

	\textit{Notation}: The complex conjugate of the variable $c$ is denoted by $c^{\ast}$. Bold lowercase letters denote column vectors and bold uppercase letters denote matrices, respectively. For the vector $\bm{v}$, $\|\bm{v}\|$ is its Euclidean norm; $\mathrm{Diag}(\bm{v})$ constructs a diagonal matrix from $\bm{v}$. The matrix transpose of $\bm{M}$ is $\bm{M}^{\mathrm{T}}$; $\bm{M}^{\mathrm{H}}$ is the Hermitian; $\mathrm{tr}(\bm{M})$ is the trace of the matrix, and $\vert \bm{M} \vert$ denotes the determinant. The notation $\bm{M}[m,n]$ refers to the element $(m,n)$ of the matrix; $\bm{I}$ denotes the identity matrix. For the random variable $\bm{x}$, $p(\bm{x}) = \text{CN}(\bm{x} \vert \bm{\mu},\bm{C})$ denotes a complex Gaussian \ac{pdf} with mean $\bm{\mu}$ and covariance matrix $\bm{C}$; $p(x) = \text{Ga}(x \vert a,b)$ is a Gamma \ac{pdf} with shape $a$ and rate parameters $b$. The expectation of the probability distribution $p(x)$ with respect to $q(x)$ (denoted in shorthand as $p$ and $q$) is written as $\expect{p}{q} = \int{p(x)q(x)\mathrm{d}x}$. The vertical bar denotes a conditional \ac{pdf} $p(\bm{x} \vert \bm{z})$; the semicolon a parametrized \ac{pdf} $p(\bm{x} ; \bm{\theta})$. $\tilde{\bm{\theta}}$ denotes the vector containing the ``true'' parameters of the \textit{generative} model and $\hat{\bm{\theta}}$ denotes an \textit{estimate} of the parameter vector $\bm{\theta}$.

\section{Signal Model}\label{sec:signal-model}

\subsection{FMCW Radar under Interference}

\begin{figure*}[ht]
\begin{center}
\scalebox{1}{\hspace{-1cm}\includegraphics{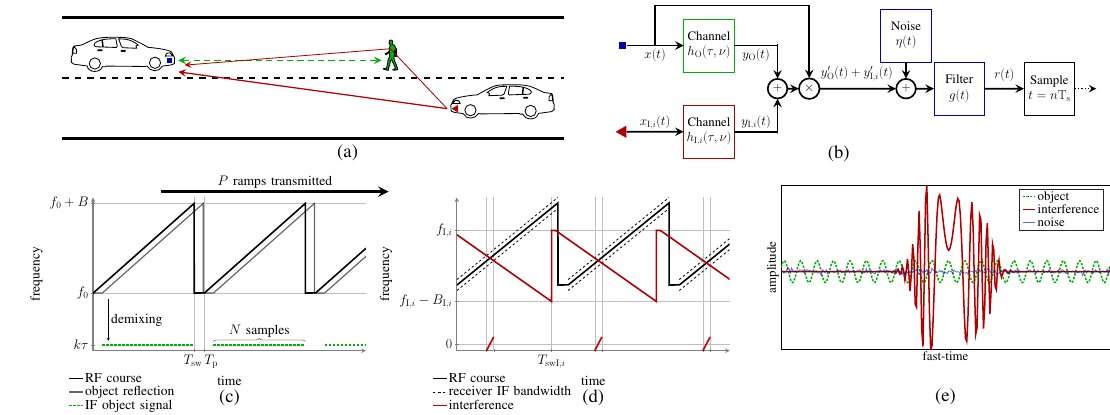}}
\caption{Illustrations of the principles and signal modeling for a mutual automotive FMCW radar interference scenario. (a) A critical automotive radar scenario with a pedestrian to be detected and an interfering vehicle on the opposite lane. (b) Abstracted block diagram representing the applied signal model. (c)-(d) Time-frequency plots of the involved signals. (e) Plot of the received signal components from an interfered ramp.}
\label{fig:interference_illustration}
\end{center}
\end{figure*}

The \textit{victim} radar transmits a transmit signal $x(t)$ in an environment containing $M_{\mathrm{I}}$ \textit{interferers} with respective transmit signals $x_{\mathrm{I},i}(t),\ist i \in [0,M_{\mathrm{I}}-1]$. The corresponding received \ac{if} signal $r(n,p)$, where $n \in [0,N-1]$ denotes the \textit{fast-time} index with sampling time $\mathrm{T_s}$ and $p \in [0,P-1]$ is the \textit{slow-time} ramp index\footnote{We consider a single receive antenna of the radar system. The proposed model and algorithm are extendable to coherent multi-antenna processing, but further investigation is out of scope of the current work.}, is given by
\begin{align}
r(t^{\prime}=n\mathrm{T_s},p) = r_{\mathrm{O}}(n,p) + \sum^{M_{\mathrm{I}}-1}_{i=0} r_{\mathrm{I},i}(n,p) + \eta(n,p)
\label{eq:IF_signal_full} \ist.\\[-7mm] \nonumber
\end{align}
The first term $r_{\mathrm{O}}(n,p)$ refers to the coherently received radar echo, termed the \textit{object signal}. The non-coherent received \textit{interference signal} from the $i$-th interferer is denoted by $r_{\mathrm{I},i}(n,p)$. Finally, $\eta(n,p)$ represents \ac{awgn} with \textit{precision} (inverse variance) $\tilde{\lambda}$. Fig.~\ref{fig:interference_illustration}a illustrates an example interference scenario where $M_\mathrm{I} = 1$, and Fig.~\ref{fig:interference_illustration}e shows example signals for a single ramp.

The victim radar, a monostatic FMCW radar, transmits a sequence of linearly modulated \textit{chirp} signals \textit{(ramps)}
\begin{align}
	&x(t) = A \sum^{P-1}_{p=0} \bar{x}(t-T_{\mathrm{p}} p; T_\mathrm{sw}) u(t-T_{\mathrm{p}} p ; f_0 , k)\label{eq:victim_transmit_signal}\\[-7mm]\nn
\end{align}
with
\begin{align}
	\bar{x}(t;T_\mathrm{sw}) &= 
		\begin{cases}
			1 & \text{for } 0 \leq t \leq T_\mathrm{sw} \\
			0 & \text{otherwise}
		\end{cases} \\
	u(t ; f_0 , k) &= \exp\bigl(\mathrm{j}(2\pi f_0 t + \pi k t^2)\bigr)
\end{align}
where we defined the parametrized rectangular function template $\bar{x}(t;T_\mathrm{sw})$ and the chirp $u(t; f_0 , k)$. $A$ is some complex transmit amplitude, $P$ is the number of chirps, $f_0$ is the chirp start frequency, $k$ is the chirp slope and $T_\mathrm{p}$ is the pulse duration which consists of the actual transmit chirp duration $T_{\mathrm{sw}}$ plus an additional idle time.\footnote{Note that actual transmit signals are real-valued but a complex receiver architecture is assumed here. Thus, we immediately use a complex representation for mathematical simplicity. However, it is possible to model a real receiver architecture as well.}

The transmit signal $x(t)$ in \eqref{eq:victim_transmit_signal} propagates through a multipath channel, yielding a superposition of reflected signals that are coherently  demodulated by multiplication with $x^{\ast}(t)$, and filtered by the receiver \ac{aaf} $G(f)$. This is indicated in Fig.~\ref{fig:interference_illustration}b starting at the rectangle symbol. The according received object signal $r_{\mathrm{O}}(t^{\prime},p)$ is given by
\vspace*{-1mm}
\begin{align}
r_{\mathrm{O}}(t^{\prime},p) &\approx H_{\mathrm{O}}(f_0 + k t^{\prime},T_{\mathrm{p}}p) \nonumber \\
&= \sum^{\tilde{L}-1}_{l=0} \tilde{\alpha}_l \exp{(-\mathrm{j} 2\pi (f_0 + k t^{\prime}) \tilde{\tau}_l )} \exp{(-\mathrm{j}2\pi \tilde{\nu}_l T_\mathrm{p}p)}
\label{eq:IF_signal_victim}\\[-7mm]\nonumber
\end{align}
with $H_{\mathrm{O}}(f,t)$ being the frequency-selective time-variant \textit{channel transfer function} of the \textit{object channel} made up of $\tilde{L}$ components\footnote{The object channel describes the full radar measurement scene, where components originate from objects in the scene. Note that a single physical object can potentially generate multiple components.} with respective amplitudes $\tilde{\alpha}_l$. A detailed derivation of \eqref{eq:IF_signal_victim} can be found in Appendix~\ref{sec:appendix:interference_model}. An illustration of a single object component in the time-frequency plane can also be seen in Fig.~\ref{fig:interference_illustration}c. The fast-time dimension with $t^{\prime} = t-T_{\mathrm{p}}p \in (0,T_{\mathrm{sw}}) \forall p$ corresponds to the \textit{beat frequencies} or analogously the \textit{delays} $\tilde{\tau}_l$ proportional to the distances of the object components. The slow-time dimension with $p \in [0,P-1]$ corresponds to the \textit{Doppler frequencies} $\tilde{\nu}_l$ proportional to object velocities. Sampling leads to the discretized fast-time domain with $t^{\prime} = n\mathrm{T_s}$.

The transmit signal of the $i$-th interferer is given by
\vspace*{-1mm}
\begin{align}
x_{\mathrm{I},i}(t) = A_{\mathrm{I},i} \sum^{P_{\mathrm{I},i}-1}_{p_{\mathrm{I}}=0} & \bar{x}(t-\bar{T}_i-T_{\mathrm{pI},i}\,p_{\mathrm{I}}; T_{\mathrm{swI},i}) \nn \\
& \times u(t-T_{\mathrm{pI},i}\,p_{\mathrm{I}} ; f_{\mathrm{I},i} , k_{\mathrm{I},i})
\label{eq:transmit_sig_int}
\end{align}
where $\bar{T}_i$ is some time offset between victim and interferer transmit sequences and other parameters are analogous to~\eqref{eq:victim_transmit_signal}. Each interferer transmit signal propagates through a multipath channel. Within our formulation, considerations for this \textit{interference channel} are analogous to the object channel. However, due to non-coherent demodulation the interferer transmit chirp sequence does not vanish, and the effect of the \ac{aaf} cannot be neglected. The model is represented in Fig.~\ref{fig:interference_illustration}b starting at the triangle symbol, and a time-frequency schematic is found in Fig.~\ref{fig:interference_illustration}d. The result reads
\vspace*{-1mm}
\begin{align}\label{eq:IF_signal_interference}
r_{\mathrm{I},i}(t^{\prime},p) &\approx \sum^{P_{\mathrm{I},i}-1}_{p_\mathrm{I} = 0} \bar{u}_{\mathrm{pI},i}(t^{\prime},p) H_{\mathrm{I},i}(\bar{f}_{\mathrm{I},i}+k_{\mathrm{I},i} t^{\prime}, T_{\mathrm{pI},i}\,p_\mathrm{I}) \nn\\ 
&\hspace*{10mm}\times \bar{G}(\bar{\Delta} f_{0,i}+{\Delta k_i}t^{\prime})\\[-7mm]\nn
\end{align}
where $\bar{u}_{\mathrm{pI},i}(t^{\prime},p)$ is the noncoherently demodulated interferer transmit chirp, and $H_{\mathrm{I},i}(f,t)$ and $\bar{G}(f)$ denote the interference channel and a modified \ac{aaf} transfer function, respectively. For derivation details, see Appendix~\ref{sec:appendix:interference_model}.

\subsection{Inference Problem Statement}

The objective is to jointly estimate the object parameters, including the number of objects (see \eqref{eq:IF_signal_victim}), as well as the interfering signal parameters and the interfering channel (see \eqref{eq:IF_signal_interference}). Using the assumptions given in Appendix~\ref{sec:appendix:interference_model}, the generic model in \eqref{eq:IF_signal_full} can be rewritten as
\vspace*{-1mm}
\begin{align}
{\hspace{-0.25cm}} r(n,p) &= \sum^{\tilde{L}-1}_{l = 0} \tilde{\alpha}_l \exp(-\mathrm{j}2\pi \tilde{\phi}_l \mathrm{T_s} n) \exp(-\mathrm{j}2\pi \tilde{\nu}_l T_{\mathrm{p}} p) \nn\\
&{\hspace{0.5cm}} + u(\mathrm{T_s} n;\Delta\tilde{f}^{(p)}_0,\Delta\tilde{k}^{(p)})\bar{G}(\Delta\tilde{f}^{(p)}_0 + \Delta\tilde{k}^{(p)} \mathrm{T_s} n) \nn\\ 
&{\hspace{0.5cm}} \times \sum^{\tilde{K}^{(p)}-1}_{k = 0} \tilde{\beta}^{(p)}_{k} \exp(-\mathrm{j}2\pi \tilde{\vartheta}^{(p)}_{k} \mathrm{T_s} n) + \eta(n,p) \,.
\label{eq:r_line}\\[-7mm]\nonumber
\end{align}
The object signal described in \eqref{eq:IF_signal_victim} is simplified, noting that the object channel component beat frequency is equal to $\tilde{\phi} = k \tilde{\tau}$. The interference signal is recast with separate parameters for each ramp $p$. For each ramp $p$, the delay-only interference channel is assumed to consist of $\tilde{K}^{(p)}$ components with respective amplitudes $\tilde{\beta}^{(p)}$ and beat frequencies $\tilde{\vartheta}^{(p)}$. Note that~\eqref{eq:r_line} constitutes a superposition of ``source signals'', each represented by a linear combination of a limited number of components, corrupted by \ac{awgn}. Each of these source signals is traditionally referred to as a \textit{line spectrum}. Inferring the underlying model parameters constitutes an instance of a \ac{lse} problem. The inference model in accordance to \eqref{eq:r_line} is given as
\begin{equation}\label{eq:inference_model_radar}
\bm{r} = \bm{\Phi}(\bm{\zeta})\bm{\alpha} + \bm{U}(\bm{\theta})\bm{\Psi}(\bm{\vartheta})\bm{\beta} + \bm{\eta}
\end{equation}
where the vector $\bm{r} \in \mathbb{C}^{PN \times 1}$ is constructed from stacking the samples of $r(p,n)$ and $p(\bm{\eta}) = \text{CN}(\bm{\eta} \vert 0,\lambda^{-1}\bm{I})$. This is an extension of the well-known model from compressed sensing~\cite{chen2001SIAMRev} and also described for \ac{mca}~\cite{starck_redundant_2004}. $\bm{\Phi}(\bm{\zeta}) = [\bm{\phi}(\bm{\zeta}_0) \iist \cdots \iist \bm{\phi}(\bm{\zeta}_{L-1})]$ is an $M \times L$ dictionary matrix with $M \leq L$, typically $M \ll L$. Note that the non-linear parameters in $\bm{\zeta}$, which define the dictionary basis vectors, may be fixed \textit{on-grid} or considered unknown and adaptively estimated, referred to as \textit{grid-less} from here on.

The object signal dictionary $\bm{\Phi}(\bm{\zeta})$ consists of the basis vectors
\begin{multline}
\bm{\phi}(\bm{\zeta}_{l} = [\varphi_l \mathrm{T_s} \,\, \nu_l T_\mathrm{p}]) = \frac{1}{\sqrt{PN}} (\exp{(-\mathrm{j}2\pi\nu_k T_\mathrm{p} \bm{p})} \\ \otimes \exp{(-\mathrm{j}2\pi\varphi_l \mathrm{T_s} \bm{n})})
\end{multline}
where $\otimes$ denotes the Kronecker product. We estimate the \textit{normalized} beat frequencies $\varphi_l \mathrm{T_s} \in [-1/2,1/2)$ and normalized Doppler frequencies $\nu_l T_\mathrm{p} \in [-1/2,1/2)$.

As the interference $\bm{U}(\bm{\theta})\bm{\Psi}(\bm{\vartheta})\bm{\beta}$ is modeled for every $p$ separately, $\bm{U}(\bm{\theta}) \in \mathbb{C}^{PN \times PN}$ and $\bm{\Psi}(\bm{\vartheta}) \in \mathbb{C}^{PN \times PK}$ with $\bm{\theta} = [\bm{\theta}^{0} \iist\cdots\iist \bm{\theta}^{P-1}]$ and $\bm{\vartheta} = [\bm{\vartheta}^{0} \iist\cdots\iist \bm{\vartheta}^{P-1}]$ are \textit{block-diagonal} matrices. I.e.,
\begin{equation}
\bm{U}(\bm{\theta}) = \begin{bmatrix}
\bm{U}(\bm{\theta}^{(p=0)}) & & \bm{0} \\
& \ddots &  \\
\bm{0} & & \bm{U}(\bm{\theta}^{(P-1)})
\end{bmatrix}
\end{equation}
with $\bm{\theta}^{(p)} = [\Delta f^{(p)}_0\,\,\Delta k^{(p)}]$ and
\begin{multline}\label{eq:inference_model_U_block}
\bm{U}(\bm{\theta}^{(p)}) = \mathrm{Diag}\bigl(u(\mathrm{T_s} \bm{n};\Delta f^{(p)}_0,\Delta k^{(p)}) \\ \times \bar{G}(\Delta f^{(p)}_0 + \Delta k^{(p)} \mathrm{T_s} \bm{n})\bigr)
\end{multline}
is an $N \times N$ diagonal matrix that contains the effect of the non-coherent demodulation and the \ac{aaf}. The channel dictionary $\bm{\Psi}(\bm{\vartheta})$ is constructed analogously from blocks $\bm{\Psi}(\bm{\vartheta}^{(p)})$ of size $N \times K$, and on-grid basis vectors $\bm{\psi}(\vartheta_k) = (1/\sqrt{N}) \exp(-\mathrm{j}2\pi\vartheta_k \mathrm{T_s} \bm{n})$, where $K \geq N$ is the chosen size of the grid. As the basis vectors are fixed, we do not write out the parameters $\bm{\vartheta}$ in the sequel.

The ill-posed problem of estimating the \textit{weights} $\bm{\alpha}$ and $\bm{\beta}$ and the according \textit{parameters} $\bm{\zeta}$ and $\bm{\theta}$ may be solvable under the assumption of sparsity, stemming from $\tilde{L}$ and $\tilde{K}$ being small in~\eqref{eq:r_line}. That is, we have some optimization problem of the form
\begin{multline}\label{eq:optimization_generic}
\hat{\bm{\alpha}},\hat{\bm{\beta}},\hat{\bm{\zeta}},\hat{\bm{\theta}} = \argmin_{\bm{\alpha},\bm{\beta},\bm{\zeta},\bm{\theta}} {\|\bm{r} - \bm{\Phi}(\bm{\zeta})\bm{\alpha} - \bm{U}(\bm{\theta})\bm{\Psi}\bm{\beta}\|}^2 \\ - f(\bm{\alpha},\bm{\beta},\bm{\zeta},\bm{\theta})
\end{multline}
with $f(\bm{\alpha},\bm{\beta},\bm{\zeta},\bm{\theta})$ being a sparsity-inducing penalty function.
For a single-source model, cost functions of this form and computationally feasible solution algorithms can be derived in a probabilistic framework~\cite{wipf2011TIP}. In particular, \ac{sbl}~\cite{tipping2003WAIS:FastMarginalSparseBayesian} has been developed for \ac{lse}. Variational formulations~\cite{tzikas_variational_2008} of \ac{sbl} have been further developed for on-grid~\cite{shutin2011TSP:fastVSBL}, as well as grid-less models~\cite{shutin2013:VSBL, badiu2017TSP:VSBL} that enable \textit{super-resolution} estimation. The \ac{sbl}-inspired framework is flexible, lending itself to extensions such as \textit{structured} line spectra~\cite{moderl_variational_2023} and \textit{dense multipath} channel models~\cite{grebien_super-resolution_2024}. Our proposed algorithm extends the concept to a superposition of line spectra from different source signals.

\subsection{Probabilistic Modeling}\label{sec:signal-model:probabilistic}

Our probabilistic model has the weights $\bm{\alpha}$ and $\bm{\beta}$ and the noise sensitivity $\lambda$ as latent variables. We apply a Gamma-Gaussian hierarchical model introducing the additional variables $\bm{\gamma}_{\alpha}$ and $\bm{\gamma}_{\beta}$, which leads to a sparse estimate~\cite{wipf2011TIP}. The non-linear parameters $\bm{\zeta}$ and $\bm{\theta}$ are considered as unknown parameters. Hence, the joint \ac{pdf} reads
\vspace*{-1mm}
\begin{multline}\label{eq:jointpdf_general}
p{(\bm{r},\bm{\alpha},\bm{\gamma}_{\alpha},\bm{\beta},\bm{\gamma}_{\beta},\lambda ; \bm{\zeta},\bm{\theta})} \\ = p{(\bm{r} \vert \bm{\alpha}, \bm{\beta}, \lambda ; \bm{\zeta},\bm{\theta})} p{(\bm{\alpha} \vert \bm{\gamma}_{\alpha})} p{(\bm{\gamma}_{\alpha})} \\ \times p{(\bm{\beta} \vert \bm{\gamma}_{\beta})} p{(\bm{\gamma}_{\beta})} p{(\lambda)}
\end{multline}

The explicit forms of the terms in~\eqref{eq:jointpdf_general} are
\begin{align}
p{(\bm{r} \vert \bm{\alpha}, \bm{\beta}, \lambda ; \bm{\zeta}, \bm{\theta})} &= \text{CN}(\bm{r} \vert \bm{\Phi}{(\bm{\zeta})}\bm{\alpha}+\bm{U}{(\bm{\theta})}\bm{\Psi}\bm{\beta},\lambda^{-1}\bm{I}) \label{eq:likelihood_dist}\\
p{(\bm{\alpha} \vert \bm{\gamma}_{\alpha})} p{(\bm{\gamma}_{\alpha})} &= \prod_l \text{CN}(\bm{\alpha}_{l} \vert 0, \gamma^{-1}_{\alpha,l}) \notag \\ 
&\qquad \quad \times \text{Ga}(\gamma_{\alpha,l} \vert a_{0},b_{0}) \label{eq:alpha_dist}\\
p{(\bm{\beta} \vert \bm{\gamma}_{\beta})} p{(\bm{\gamma}_{\beta})} &= \prod_p \prod_k \text{CN}(\bm{\beta}^{(p)}_{k} \vert 0, {\gamma^{-1}_{\beta,k}}^{(p)}) \notag \\ 
&\qquad \quad \times \text{Ga}(\gamma^{(p)}_{\beta,k} \vert c_{0},d_{0}) \label{eq:beta_dist}\\
p(\lambda) &= \text{Ga}(\lambda \vert e_0,f_0)\,. \label{eq:lambda_dist}
\end{align}
The likelihood function $p{(\bm{r} \vert \bm{\alpha}, \bm{\beta}, \lambda ; \bm{\zeta}, \bm{\theta})}$ is described by a Gaussian distribution due to the \ac{awgn} assumption in~\eqref{eq:r_line}. The weights $\bm{\alpha}$ and $\bm{\beta}$ are modeled as conditionally independent zero-mean Gaussian-distributed with individual \textit{precisions} governed respectively by Gamma-distributed hyper-parameters $\bm{\gamma}_\alpha$ and $\bm{\gamma}_\beta$~\cite{tipping2003WAIS:FastMarginalSparseBayesian}. The Gamma distribution is the conjugate prior for the precision of a Gaussian~\cite{bishop2009} and is known to promote sparsity~\cite{wipf2011TIP}. Consequently, the number of components, $\hat{K}$ and $\hat{L}$, can be indirectly estimated by inferring the parameters, $\bm{\gamma}_\alpha$ and $\bm{\gamma}_\beta$. The prior \ac{pdf} of the noise precision $\lambda$ is also assumed to be Gamma distributed.

\section{Inference Algorithm}
\label{sec:algorithm}

\subsection{Variational Formulation}

Directly solving the high-dimensional non-linear estimation problem of~\eqref{eq:optimization_generic} by statistical inference on the probabilistic model of~\eqref{eq:jointpdf_general} is computationally infeasible. Therefore, we resort to the variational \ac{em} solution~\cite{tzikas_variational_2008} to iteratively determine all marginal \acp{pdf} of the latent variables and \ac{ml} estimates of the unknown parameters.

Denoting the inferred \textit{proxy} posterior distribution of the latent variables as $q(\bm{\alpha},\bm{\gamma}_{\alpha},\bm{\beta},\bm{\gamma}_{\beta},\lambda)$, the \ac{em} framework iteratively maximizes the \textit{functional} 
\vspace*{-1mm}
\begin{multline}\label{eq:elbo}
\mathcal{L}(q(\bm{\alpha},\bm{\gamma}_{\alpha},\bm{\beta},\bm{\gamma}_{\beta},\lambda)) \\ = {\left\langle \log \dfrac{p{(\bm{r},\bm{\alpha},\bm{\gamma}_{\alpha},\bm{\beta},\bm{\gamma}_{\beta},\lambda ; \bm{\zeta},\bm{\theta})}}{q(\bm{\alpha},\bm{\gamma}_{\alpha},\bm{\beta},\bm{\gamma}_{\beta},\lambda)} \right\rangle}_{q(\bm{\alpha},\bm{\gamma}_{\alpha},\bm{\beta},\bm{\gamma}_{\beta},\lambda)}
\end{multline}
termed the \ac{elbo}.

For fixed parameter estimates $\hat{\bm{\zeta}}$ and $\hat{\bm{\theta}}$, the \ac{elbo} is maximized if $q$ is equal to the joint posterior \ac{pdf}. As this posterior distribution is intractable, we first constrain the form of $q$ and maximize the \ac{elbo} under those constraints to obtain a tractable \textit{variational approximation}. In line with variational \ac{sbl} \cite{shutin2011TSP:fastVSBL,shutin2013:VSBL}, we use the \textit{structured mean-field} approach~\cite{tzikas_variational_2008}
\begin{align}
q(\bm{\alpha},\bm{\gamma}_{\alpha},\bm{\beta},\bm{\gamma}_{\beta},\lambda;\bm{\zeta},\bm{\theta}) &= q(\bm{\alpha}) \prod_l  q(\gamma_{\alpha,l}) \nonumber\\ &{\hspace{0.75cm}} \times q(\bm{\beta}) \prod_k  q(\gamma_{\beta,k}) q(\lambda)\,.
\label{eq:variational_posterior}
\end{align}
I.e., groups of variables are constrained to factorize in the \ac{pdf} leading to a set of factors $\mathcal{Q}$, $q = \prod_{\mathcal{Q}} q_i$, with $q_i$ being the i-th proxy distribution. To be noted is that our model does not factorize elements of the weight vectors $\bm{\alpha}$ and $\bm{\beta}$, hence posterior correlations between these elements are taken into account. However, the proxy \acp{pdf} $q(\bm{\alpha})q(\bm{\beta}) \neq q(\bm{\alpha,\beta})$, so that different source signals \emph{do} in fact factorize, which is a simplifying assumption.\footnote{If this model assumption is not applied, we essentially obtain a single larger concatenated dictionary $[\bm{\Phi}\,\,\,\bm{U}(\bm{\theta})\bm{\Psi}]$ with its appropriate weights and joint covariance matrix. This leads to a similar algorithm, but without distinct object and interference estimation subroutines. It will be evaluated in Section~\ref{subsec:fund_anal} as a point of comparison to the proposed method, but will not be further discussed in this work.}

It can be shown that the log-distribution of the $i$-th factor $q_i$, when every other variable and parameter is fixed, is computed by
\begin{align}\label{eq:mean_field_update}
\log q_i = \expect{\log p{(\bm{r},\bm{\alpha},\bm{\gamma}_{\alpha},\bm{\beta},\bm{\gamma}_{\beta},\lambda ; \hat{\bm{\zeta}},\hat{\bm{\theta}})}}{\bar{q}_i} + \text{ const.}
\end{align}
where $\bar{q}_i$ is shorthand notation for ``proxy \ac{pdf} of every factor except for the $i$-th''. Hence, we obtain a set of interdependent implicit equations that are solved by iteratively \textit{updating} the inferred \acp{pdf}. As long as the \ac{elbo} is increased at every step, the ordering of updates is in principle arbitrary. In practice though, certain update schemes are used to improve the convergence properties of the algorithm. In the proposed algorithm, we only directly use~\eqref{eq:mean_field_update} for the noise precision $q_{\lambda}$. The proxies of the weights $q_\alpha$ and $q_\beta$ and the according unknown parameters $\bm{\zeta}$ and $\bm{\theta}$ are updated jointly~\cite{badiu2017TSP:VSBL}, which is described in Section~\ref{sec:algorithm:paramestim}. For the weight precision hyper-parameter distributions $q_\gamma$ the \textit{fast} update scheme~\cite{shutin2011TSP:fastVSBL} is used, discussed in Section~\ref{sec:algorithm:fastupdate}. The complete resulting scheme is sketched as pseudo-code in Algorithm~\ref{alg:FastVar_full}, making use of Algorithms~\ref{alg:IntEst}-\ref{alg:FastUpdate} as subroutines. The details are discussed in the sequel.

\subsection{Estimation of Unknown Parameters}
\label{sec:algorithm:paramestim}

In the typical variational \ac{em} approach, the updates of unknown parameters are derived with the estimates of the proxy \acp{pdf} fixed~\cite{tzikas_variational_2008}. However, in our application estimates of the object channel parameters $\hat{\bm{\zeta}}$ are strongly tied to the inferred weight proxy \ac{pdf} $q_{\alpha}$. Similarly, the estimation of interference chirp parameters $\hat{\bm{\theta}}^{(p)}$ at ramp $p$ strongly depends on the respective channel weights \ac{pdf} $q_{\beta}^{(p)}$. Term-wise optimization of the \ac{elbo} therefore leads to slower convergence as well as the algorithm being more prone to reach local optima. To mitigate this, the \ac{elbo} is maximized jointly for $\hat{\bm{\zeta}}$ and $q_{\alpha}$ as well as $\hat{\bm{\theta}}^{(p)}$ and $q_{\beta}^{(p)}$ for all $p$. Rewriting~\eqref{eq:elbo}, this yields the same update equations as~\eqref{eq:mean_field_update} for the resulting proxy \acp{pdf}. For the parameters, however, we obtain
\vspace*{-1mm}
\begin{multline}\label{eq:joint_param_update}
{\hspace{-0.25cm}} \hat{\bm{\theta}}^{(p)} = \argmax_{\bm{\theta}^{(p)}} \\ {\hspace{0.5cm}} {\int \exp \expect{\log p{(\bm{r},\bm{\alpha},\bm{\gamma}_{\alpha},\bm{\beta},\bm{\gamma}_{\beta},\lambda ; \hat{\bm{\zeta}},\hat{\bm{\theta}})}}{\bar{q}_{\beta}^{(p)}} \mathrm{d}\bm{\beta}^{(p)}}
\end{multline}
for the chirp parameters $\hat{\bm{\theta}}^{(p)}$. The result for the object channel dispersion parameter estimates $\hat{\bm{\zeta}}$ is analogous, where estimates for the individual components $\hat{\bm{\zeta}}_l$ can be computed separately~\cite{badiu2017TSP:VSBL,hansen_superfast_2018} for each spectral line $l$. More details on the derivation are found in Appendix~\ref{sec:appendix:fast_updates}.

\subsection{Fast Component Precision Update and Thresholding}
\label{sec:algorithm:fastupdate}

In~\cite{shutin2011TSP:fastVSBL} it has been shown for variational \ac{sbl} that with the other variables fixed, the estimates of the component precision hyper-parameters ``at infinity'' can be derived analytically, yielding a test for component convergence. Hence, at every iteration the dictionary may be adaptively pruned of divergent components, or possibly new components may be added. With this, algorithm convergence is much accelerated. Crucially, it can be seen that for our proposed extended model, the update equations for the factors corresponding to each source signal are identical to the single-source case except with $\bm{r}$ exchanged by the respective ``residuals'' $\hat{\bm{r}}_{\alpha}$ and $\hat{\bm{r}}_{\beta}$. This means that the formulations of~\cite{shutin2011TSP:fastVSBL, shutin2013:VSBL} can be followed; Appendix~\ref{sec:appendix:fast_updates} contains more details.

Using the estimates of the object signal parameters $\hat{\bm{\Phi}}$, $\hat{\bm{\alpha}}$, and $\hat{\bm{\gamma}}_{\alpha}$, the update equation, the pruning of existing components, and the addition of new potential components (see Algorithm~\ref{alg:FastUpdate}) are based on
\begin{align}
\rho_l &= {(\hat{\lambda} \hat{\bm{\phi}}^{\mathrm{H}}_l \hat{\bm{\phi}}_l - {\hat{\lambda}}^2 \hat{\bm{\phi}}^{\mathrm{H}}_l \hat{\bm{\Phi}}_{\bar{l}} \hat{\bm{C}}_{\alpha,\bar{l}} \hat{\bm{\Phi}}^{\mathrm{H}}_{\bar{l}} \hat{\bm{\phi}}_l)}^{-1} \label{eq:fast_rho} \\
\omega^2_l &= {\vert (\hat{\lambda} \rho_l \hat{\bm{\phi}}^{\mathrm{H}}_l \hat{\bm{r}}_{\alpha} - {\hat{\lambda}}^2 \rho_l \hat{\bm{\phi}}^{\mathrm{H}}_l  \hat{\bm{\Phi}}_{\bar{l}} \hat{\bm{C}}_{\alpha,\bar{l}} \hat{\bm{\Phi}}^{\mathrm{H}}_{\bar{l}} \hat{\bm{r}}_{\alpha}) \vert}^{2} \label{eq:fast_omega} \ist.
\end{align}
The notation of the subscript $l$ means taking the $l$-th element or column corresponding to that component, and $\bar{l}$ denotes the vector or matrix with said element removed or computed \textit{as if} it were removed. Existing components or new components are kept \textit{if and only if} $\rho_l/\omega_l^2 > T$ with $T=1$, where $T$ denotes a threshold and $\omega^2_l / \rho_l -1$ can be interpreted as an estimated \textit{component \ac{snr}}. Note that the original convergence test threshold given by $T=1$ leads to positive bias in the number of estimated components \cite{shutin2011TSP:fastVSBL,grebien_super-resolution_2024}. In~\cite{leitinger2020Asilomar, grebien_super-resolution_2024}, the authors propose an adaptation of the threshold $T$ for grid-less \ac{sbl}-type methods based on extreme value analysis.

Similar update equations are found for the interference parameters $\hat{\bm{\theta}}^{(p)}$, $\hat{\bm{\Psi}}^{(p)}$, $\hat{\bm{\beta}}^{(p)}$, and $\hat{\bm{\gamma}}_{\beta}^{(p)}$ for all $p$ (see Algorithm~\ref{alg:IntEst}).

\subsection{Algorithm Implementation}
\label{sec:algorithm:implementation}

We have now derived structured mean-field variational \ac{em} update steps for inference of our model. As mentioned, these steps are interdependent and therefore to be applied iteratively. Explicit forms of the update equations are listed in Appendix~\ref{sec:appendix:fast_updates}. The algorithm converges to an optimum of the \ac{elbo} corresponding to an approximate solution of our signal separation problem. However, it does not necessarily converge to the \textit{global} optimum, nor is any \textit{rate} of convergence guaranteed. In addition to the specific update schemes presented previously, the overall \textit{scheduling} of steps as well as the \textit{initialization} significantly influence the behavior of the algorithm.

\begin{algorithm}[ht]
\caption{Variational Signal Separation \textbf{(Main)}}\label{alg:FastVar_full}
\begin{algorithmic}
\Require Received signal $\bm{r}$; thresholds $T_\alpha$, $T_\beta$
\Statex \Comment{\textit{Bottom-up} dictionary initialization:}
\State $\hat{\bm{\Phi}} \gets [\,]$, $\hat{\bm{\gamma}}_{\alpha} \gets [\,]$, $\hat{\bm{\alpha}} \gets [\,]$, $\hat{\bm{C}}_{\alpha} \gets [\,]$
\State $\hat{\bm{\Psi}} \gets [\,]$, $\hat{\bm{\gamma}}_{\beta} \gets [\,]$, $\hat{\bm{\beta}} \gets [\,]$, $\hat{\bm{C}}_{\beta} \gets [\,]$
\Statex \Comment{Noise precision initialization:}
\State $\hat{\lambda} \gets (2(NP-1))/{\|\bm{r}\|}^2$
\Repeat \Comment{Main iteration}
\Statex \Comment{Update interference estimate:}
\ForAll{$p$} \Comment{per-ramp update}
\State $\hat{\bm{\theta}}^{(p)} \gets $ from~\eqref{eq:update_theta_p}
\State \textbf{run} Algorithm~\ref{alg:IntEst}($\hat{\bm{r}}^{(p)}_{\beta}$,$\bm{U}^{(p)}(\hat{\bm{\theta}})$,$\hat{\bm{\Psi}}^{(p)}$,$\hat{\bm{C}}^{(p)}_{\beta}$,$\hat{\bm{\gamma}}^{(p)}_{\beta}$,$T_{\beta}$) 
\EndFor
\State $\hat{\lambda} \gets $ from~\eqref{eq:update_lambda_hat} \Comment{update noise precision estimate}
\State $\hat{\bm{r}}_{\alpha} = \bm{r}-\bm{U}(\hat{\bm{\theta}})\hat{\bm{\Psi}}\hat{\bm{\beta}}$ \Comment{update object signal residual}
\State $\hat{\bm{C}}_{\alpha} \gets $ from~\eqref{eq:update_C_alpha} \Comment{update with new residual}
\State $\hat{\bm{\alpha}} \gets $ from~\eqref{eq:update_alpha_hat}
\Statex \Comment{Update object signal estimate:}
\State \textbf{run} Algorithm~\ref{alg:ObjEst}($\hat{\bm{r}}_{\alpha}$,$\hat{\bm{\Phi}}$,$\hat{\bm{C}}_{\alpha}$,$\hat{\bm{\gamma}}_{\alpha}$,$T_{\alpha}$)
\State $\hat{\lambda} \gets $ from~\eqref{eq:update_lambda_hat}
\State $\hat{\bm{r}}_{\beta} = \bm{r}-\hat{\bm{\Phi}}\hat{\bm{\alpha}}$
\State $\hat{\bm{C}}_{\beta} \gets $ from~\eqref{eq:update_C_beta}
\State $\hat{\bm{\beta}} \gets $ from~\eqref{eq:update_beta_hat}
\Until{convergence} \Comment{criterion or fixed number}
\end{algorithmic}
\end{algorithm}
\vspace*{-2mm}

Algorithm~\ref{alg:FastVar_full} describes the implementation of the proposed iterative algorithm. For notational simplicity, we consistently omit iteration counters; within the sequence of update steps, the algorithm uses the most recent values of other estimates. The proposed algorithm structure envisions a \textit{bottom-up} scheme, i.e., that the dictionaries $\hat{\bm{\Phi}}$ and analogously $\hat{\bm{\Psi}}$ are initialized as empty. The noise precision is initialized as $\hat{\lambda} = (2(NP-1))/{\|\bm{r}\|}^2$, i.e., half the received signal power is assumed to be from noise initially.

The main iteration consists of two separated subroutines for the estimations of the respective object and interference signals, given in Algorithms~\ref{alg:IntEst} and~\ref{alg:ObjEst}. They are similar in structure and connected by computations of $\hat{\bm{r}}_{\alpha}$, $\hat{\bm{r}}_{\beta}$, as well as $\hat{\lambda}$. I.e., the estimated residuals are used at every iteration as inputs to the respective subroutines. This scheme is obtained as the interference and object signals are additive and their components factorize according to~\eqref{eq:jointpdf_general} and~\eqref{eq:variational_posterior}. Furthermore, as interference is modeled independently over the ramps, separate steps over the index $p$ are applied. The estimation of multipath channels is based on the repeated addition and then update of line spectral components, in accordance with the bottom-up scheme, by applying~\eqref{eq:joint_param_update} and Algorithm~\ref{alg:FastUpdate}. The routine adds components it can ``find'' given the selected threshold, which are then refined or pruned using the now updated estimate of the weight covariance matrix.

\vspace*{-2mm}
\begin{algorithm}[ht]
\caption{Interference Signal Estimation \textbf{(Subroutine)}}\label{alg:IntEst}
\begin{algorithmic}
\Require Estimates $\hat{\bm{r}}$, $\hat{\bm{\theta}}$, $\hat{\bm{\Psi}}$, $\hat{\bm{C}}$, $\hat{\lambda}$, $\hat{\bm{\gamma}}$; threshold $T$
\State $\hat{k} \gets \argmax_{\bar{k}} {({\omega^2}_{\bar{k}} / \rho_{\bar{k}})}$ where $\bar{k} \in \bar{\mathcal{K}}$ \\ \Comment{find possible new component}
\State \textbf{run} Algorithm~\ref{alg:FastUpdate}($\hat{\bm{r}}$,$\bm{U}(\hat{\bm{\theta}})\hat{\bm{\Psi}}$,$\hat{\bm{C}}$,$\hat{\bm{\gamma}}$,$\hat{k}$,$\bm{U}(\hat{\bm{\theta}})\hat{\bm{\psi}}_{\hat{k}}$,$T$) \\ \Comment{add component if above threshold}
\ForAll{$k \in \mathcal{K}$} \Comment{update all existing components}
\State \textbf{run} Algorithm~\ref{alg:FastUpdate}($\hat{\bm{r}}$,$\bm{U}(\hat{\bm{\theta}})\hat{\bm{\Psi}}$,$\hat{\bm{C}}$,$\hat{\bm{\gamma}}$,$k$,$\hat{\bm{\psi}}_k$,$T$)
\EndFor
\State $\hat{\bm{\beta}} \gets $ from~\eqref{eq:update_beta_hat}
\State $\hat{\bm{\theta}} \gets $ from~\eqref{eq:update_theta_p} \Comment{update chirp parameter estimates}
\end{algorithmic}
\end{algorithm}
\vspace*{-2mm}

\begin{algorithm}[ht]
\caption{Object Signal Estimation \textbf{(Subroutine)}}\label{alg:ObjEst}
\begin{algorithmic}
\Require Estimates $\hat{\bm{r}}$, $\hat{\bm{\Phi}}$, $\hat{\bm{C}}$, $\hat{\lambda}$, $\hat{\bm{\gamma}}$; threshold $T$
\State $\hat{l} \gets L$ \Comment{possible new component index}
\State $\hat{\bm{\zeta}}_{\hat{l}} \gets $ from~\eqref{eq:update_zeta_l}
\State $\hat{\bm{\phi}}_{\hat{l}} \gets $ generate according to signal model
\State \textbf{run} Algorithm~\ref{alg:FastUpdate}($\hat{\bm{r}}$,$\hat{\bm{\Phi}}$,$\hat{\bm{C}}$,$\hat{\bm{\gamma}}$,$\hat{l}$,$\hat{\bm{\phi}}_{\hat{l}}$,$T$) \\ \Comment{add component if above threshold}
\State $\hat{\bm{\alpha}} \gets $ from~\eqref{eq:update_alpha_hat}
\ForAll{$l$} \Comment{update all existing components}
\State $\hat{\bm{\zeta}}_{l} \gets $ from~\eqref{eq:update_zeta_l}
\State $\hat{\bm{\phi}}_{l} \gets $ update with new $\hat{\bm{\zeta}}_{l}$
\State \textbf{run} Algorithm~\ref{alg:FastUpdate}($\hat{\bm{r}}$,$\hat{\bm{\Phi}}$,$\hat{\bm{C}}$,$\hat{\bm{\gamma}}$,$l$,$\hat{\bm{\phi}}_l$,$T$)
\State $\hat{\bm{\alpha}} \gets $ from~\eqref{eq:update_alpha_hat}
\EndFor
\end{algorithmic}
\end{algorithm}
\vspace*{-1mm}

\begin{algorithm}[ht]
\caption{Fast Component Precision Update \textbf{(Subroutine)}}\label{alg:FastUpdate}
\begin{algorithmic}
\Require Estimates $\hat{\bm{r}}$, $\hat{\bm{\Phi}}$, $\hat{\bm{C}}$, $\hat{\lambda}$, $\hat{\bm{\gamma}}$; existing or new component index $l$, basis vector $\hat{\bm{\phi}}_l$, threshold $T$
\State $\rho_l \gets $ from~\eqref{eq:fast_rho}
\State $\omega^2_l \gets $ from~\eqref{eq:fast_omega}
\If{$\omega^2_l / \rho_l > T$}
\State $\hat{\gamma}_{l} \gets (1/(\omega^2_l - \zeta_l))$
\If{$l \notin \mathcal{L}$} \Comment{add new component}
\State $\hat{\bm{\gamma}} \gets (\hat{\bm{\gamma}} \cup \hat{\gamma}_{l})$
\State $\hat{\bm{\Phi}} \gets (\hat{\bm{\Phi}} \cup \hat{\bm{\phi}}_{l})$
\State $L \gets L+1$
\EndIf
\Else
\If{$l \in \mathcal{L}$} \Comment{prune existing component}
\State $\hat{\bm{\gamma}} \gets (\hat{\bm{\gamma}} \setminus \hat{\gamma}_{l})$
\State $\hat{\bm{\Phi}} \gets (\hat{\bm{\Phi}} \setminus \hat{\bm{\phi}}_l)$
\State $L \gets L-1$
\EndIf
\EndIf
\State $\hat{\bm{C}} \gets $ from~\eqref{eq:update_C_alpha} using updated $\hat{\bm{\gamma}}$ and $\hat{\bm{\Phi}}$
\end{algorithmic}
\end{algorithm}
\vspace*{-2mm}

Algorithm~\ref{alg:FastVar_full} starts by estimating the interference as described by Algorithm~\ref{alg:IntEst}. The interference channel is estimated on a grid. Hence, the possible ``complete'' set of basis vectors of the dictionary matrix is fixed, denoted by the index sets $\mathcal{K}$ and $\bar{\mathcal{K}} = \{[0,K-1] \setminus \mathcal{K}\}$ for the respective \textit{active} and \textit{passive} sets of basis vectors. At every iteration, we select the passive component with the largest estimated component \ac{snr} to potentially add to the active dictionary. Irrespective whether or not a new component was added, we proceed with updating all active components. Only a single component is potentially added per main iteration to avoid $\hat{\bm{\Psi}}$ converging before the chirp parameter estimates $\hat{\bm{\theta}}$. $\hat{\bm{\theta}}$ is updated once before and after this dictionary update process. The initial chirp parameter estimates are provided by evaluating the cost function on a coarse grid. Further estimates are obtained by applying any established \textit{constrained optimization} algorithm, with possible values of $\hat{\bm{\theta}}$ constrained to a reasonable range.

For the grid-less object signal estimation of Algorithm~\ref{alg:ObjEst}, the basis vectors are generated and refined adaptively according to the respective estimates $\hat{\bm{\zeta}}_l$. Also here, only at most a single component is added per iteration. In this way, the chance of converging to local optima by wrongly assigning part of the interference to the object estimate is minimized. Finally, note that the value of the component threshold for the coherently received object signal $T_\alpha$ can be systematically set using the analysis in~\cite{leitinger2020Asilomar} to achieve a certain \textit{false alarm rate}. This is not the case for its interference counterpart $T_\beta$ due to the unknown $\bm{\theta}$. As the object signal is coherently processed over all ramps, $T_\alpha > T_\beta$ is reasonable. Empirical testing has shown that a low threshold between $0 \mathrm{dB}$ and $6 \mathrm{dB}$ for the interference is generally desirable.

\section{Analysis and Results}
\label{sec:results}

\subsection{Fundamental Object Estimation Performance}
\label{subsec:fund_anal}

The main task of the radar sensor is to accurately determine the distances and velocities of surrounding objects. Therefore, we evaluate the proposed algorithm based on the quality of object parameter estimates. First, a comparison to the \ac{crlb}~\cite{kay_fundamentals_1993} on the estimation error variance is carried out. In order to simplify the analysis and numerical evaluations, a scenario with a single transmit ramp, and hence delay-only estimation, is set up. The object channel consists of a single line spectral component whereas the interference channel is constant, i.e., there is only direct-path interference corresponding to a single component at excess delay zero. The interference signal template is as described in Section~\ref{sec:signal-model} and Appendix~\ref{sec:appendix:interference_model}. The \ac{aaf} frequency transfer function is a raised cosine with Nyquist bandwidth $f_\mathrm{s}/4$ and roll-off factor 0.25. Relevant signal parameters for this experiment are in Table~\ref{tab:params} (column denoted Simulation I).
\vspace*{2mm}

\begingroup
\setlength{\tabcolsep}{2pt}
\renewcommand{\arraystretch}{1.2}
\begin{table*}
\caption{Summary of relevant signal parameters for the analysis. $\mathrm{c_0}$ denotes the speed of light in a vacuum.}
\label{tab:params}
\vspace*{-2mm}
\centering
\scalebox{.95}{\begin{tabular}{ c | c | c | c | c }
 \textbf{Parameter} & \textbf{Simulation I} & \textbf{Simulation II} & \textbf{Obj. Measurement} & \textbf{Int. Measurement} \\ 
 \hline
 \multicolumn{5}{c}{\textbf{Victim Radar and Object Channel}} \\
 \hline
 Ramp start frequency $f_0 (\mathrm{GHz})$ & $79$ & $79$ & $79$ & $76.2$ \\
 Ramp slope $k (\mathrm{GHz / s})$ & $10^{4}$ & $10^{4}$ & $2.08 \times 10^{4}$ & $2.08 \times 10^{4}$ \\
 Ramp duration $T_{\mathrm{sw}} (\mathrm{\mu s})$ & $25$ & $25$ & $6.39$ & $24$ \\
 Pulse duration $T_{\mathrm{p}} (\mathrm{\mu s})$ & $25$ & $25$ & $44$ & $44$ \\
 No. of ramps $P$ & 1 & 16 & 32 & 32 \\
 No. of samples $N$ & 256 & 128 & 256 & 256 \\
 Sampling frequency $\mathrm{f_s} = 1/\mathrm{T_s} (\mathrm{MHz})$ & 10.2 & 5.1 & 40 & 40 \\
 No. of components $\tilde{L}$ & 1 & 10 & unknown & unknown \\
 & & & & \\
 Component weights $\tilde{\alpha}$ & \makecell[cc]{$1\exp{(\mathrm{j}\varphi_\alpha)}$, \\ $\varphi_\alpha \sim \mathcal{U}(0,2\pi)$} & \makecell[cc]{$\log {\vert \alpha \vert}^{2}_{[\mathrm{dB}]}$ \\ $= -40\log(\tilde{\tau}\mathrm{c_0}+1)+x$, \\ $x \sim \mathcal{U}(-3,3)$} & unknown & unknown \\
 & & & & \\
 Component delays $\tilde{\tau} (\mathrm{ns})$ & $80.06$ & $\mathcal{U}(3.97,127)$ & unknown & unknown \\
 Component Doppler frequencies $\tilde{\nu} (\mathrm{kHz})$ & $0$ & $\mathcal{U}(-5,5)$ & unknown & unknown \\
 \hline
 \multicolumn{5}{c}{\textbf{Interferer and Interference Channel}} \\
 \hline
 Ramp start frequency $f_{\mathrm{I}} (\mathrm{GHz})$ & $79.01$ & $\{ 79.002,79.004,79.008 \}$ & - & $76.1$ \\
 Ramp slope $k_{\mathrm{I}} (\mathrm{GHz / s})$ & $9.2 \times 10^{3}$ & $\{ 9.8321,9.7122,9.3925 \}\times 10^{3}$ & - & $2.6923 \times 10^{4}$ \\
 Ramp duration $T_{\mathrm{swI}} (\mathrm{\mu s})$ & $25$ & $25.02$ & - & $26$ \\
 Pulse duration $T_{\mathrm{pI}} (\mathrm{\mu s})$ & $25$ & $\{75.01,50.01,25.02 \}$ & - & $25$ \\ 
 No. of ramps $P_{\mathrm{I}}$ & 1 & $\{2,4,8\}$ & - & $8$ \\
 No. of components $\tilde{K}$ & 1 & 10 & unknown & unknown \\
  & & & & \\
 Component weights $\tilde{\beta}$ & \makecell[cc]{$\vert \beta \vert\exp{(\mathrm{j}\varphi_\beta)}$, \\ $\vert \beta \vert$ according to $\mathrm{SIR}$, \\ $\varphi_\beta \sim \mathcal{U}(0,2\pi)$} & \makecell[cc]{$\log {\vert \beta \vert}^{2}_{[\mathrm{dB}]}$ \\ $= -20\log(\tilde{\tau}_{\mathrm{I}}\mathrm{c_0}+1)+x$, \\ $x \sim \mathcal{U}(-10,0)$} & - & unknown \\
  & & & & \\
 Component delays $\tilde{\tau}_{\mathrm{I}} (\mathrm{ns})$ & $0$ & \makecell[cc]{0 for first component, \\ $\mathcal{U}(3.97,127)$ for others} & - & unknown \\
\end{tabular}}
\end{table*}
\endgroup
\vspace*{-2mm}

\begin{figure}[ht]
\begin{center}
\scalebox{1}{\includegraphics{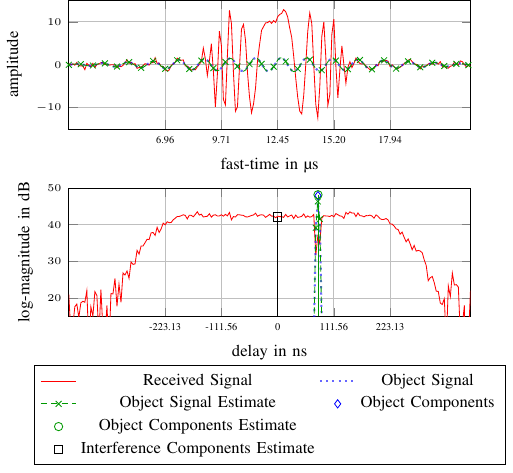}}
\caption{Example signal and object estimate at $\mathrm{SNR} = 30\mathrm{dB}$ and $\mathrm{SIR} = -15\mathrm{dB}$. To reduce visual clutter, only certain signals are shown. The upper plot contains the fast-time view of the complete received signal $\bm{r}$, the object source signal $\bm{r}_{\mathrm{O}}$ and its estimate $\hat{\bm{r}}_{\mathrm{O}}$. The lower plot contains the same signals in the delay domain, as well as the true object channel components ($\tilde{\bm{\zeta}}$ and $\tilde{\bm{\alpha}}$), their estimates ($\hat{\bm{\zeta}}$ and $\hat{\bm{\alpha}}$), and the estimated interference channel components ($\hat{\bm{\vartheta}}$ and $\hat{\bm{\beta}}$).}
\label{fig:sig_estEx}
\end{center}
\end{figure}

We study the expected \ac{rmse} of the normalized beat frequency estimate $\sqrt{\expect{\|\hat{\varphi} - \tilde{\varphi}\|^{2}}{}}$ over $500$ simulated realizations. The expectation is replaced by the sample average, where only realizations with correct model order estimates $\hat{L} = \tilde{L} = 1$ are taken into account. Different realizations consist of varying the noise as well as the phases of the complex scalar weights $\tilde{\alpha}$ and $\tilde{\beta}$. The \ac{rmse} is analyzed for different \acp{snr} and \acp{sir}, defined as
\vspace*{-1mm}
\begin{align}
\mathrm{SNR} &= \lambda \|\bm{\Phi}(\tilde{\bm{\zeta}})\tilde{\bm{\alpha}}\|^2 \label{eq:SNR}\\
\mathrm{SIR} &= \|\bm{\Phi}(\tilde{\bm{\zeta}})\tilde{\bm{\alpha}}\|^2 \, / \, \|\bm{U}(\tilde{\bm{\theta}})\tilde{\bm{\Psi}}\tilde{\bm{\beta}}\|^2 \, . \label{eq:SIR}
\end{align}
Three algorithm variants are considered. First, mutual interference is completely neglected in the inference model, i.e., Algorithm~\ref{alg:IntEst} and associated steps are removed. Second, the \textit{factorized} algorithm as proposed in this work can be applied. Third, as mentioned in Section~\ref{sec:algorithm}, it is possible to \textit{jointly} model the interference and object channels within a larger concatenated dictionary by modeling the joint weight proxy $q(\bm{\alpha},\bm{\beta})$ in~\eqref{eq:variational_posterior}. For all variants, component acceptance thresholds are set to $T_{\alpha} = 9\mathrm{dB}$ and $T_{\beta} = 3\mathrm{dB}$. Results are compared to the root mean \ac{crlb} whose values are computed from the given signal model of~\eqref{eq:inference_model_radar}. Fig.~\ref{fig:sig_estEx} shows an example result of the proposed algorithm.

\begin{figure*}[ht]
\begin{center}
\scalebox{1}{\includegraphics{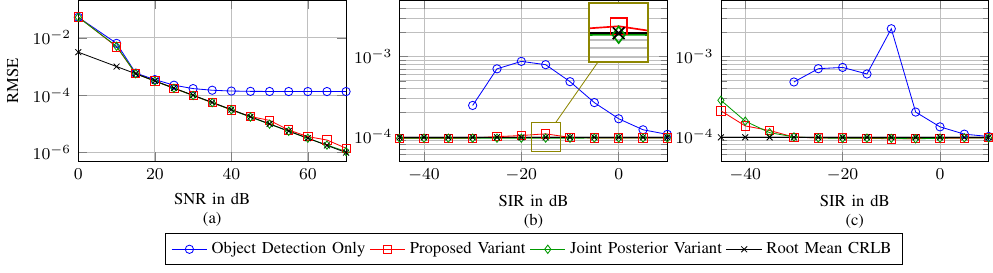}}
\caption{Statistical results of single-object delay estimation \ac{rmse}. (a) Evaluated over varying \ac{snr} at a fixed \ac{sir} of $0\mathrm{dB}$. (b) Evaluated over varying \ac{sir} at a fixed \ac{snr} of $30\mathrm{dB}$. (c) Evaluated for the \ac{aaf} model error scenario.}
\label{fig:CRLB_group}
\end{center}
\end{figure*}

In Fig.~\ref{fig:CRLB_group}a it can be seen for a fixed $\mathrm{SIR} = 0\mathrm{dB}$ that the algorithm jointly modeling object and interference channels achieves the \ac{crlb} once the \ac{snr} is high enough for the object channel component to be detected. The factorized algorithm only experiences a slight performance loss visible at high \acp{snr}, whereas the ``object only'' algorithm, as expected, cannot achieve an estimation error below a certain value determined by the disturbing interference.

We also observe the performance over varying \ac{sir} values, plotted in Fig.~\ref{fig:CRLB_group}b for a fixed $\mathrm{SNR} = 30\mathrm{dB}$. The performance of object-only channel estimation strongly deteriorates with increasing interference power. Although the \ac{rmse} seemingly recovers at lower \ac{sir}, this is misleading as the estimate gets dominated by the interference signal and the number of missed detections is growing. Below around $-30\mathrm{dB}$, the component cannot be detected at all. This is not the case for the signal separation-based methods. The proposed factorized algorithm shows its largest evaluated deviation from the theoretical optimum at $-15\mathrm{dB}$ (highlighted by a close-up). We observe that the worst case for the factorized algorithm is related to the condition
\begin{equation}\label{eq:coherence}
\| \tilde{\bm{\Phi}}^{\mathrm{H}}\tilde{\bm{\Phi}}\tilde{\bm{\alpha}} \| \approx \| \tilde{\bm{\Phi}}^{\mathrm{H}}\bm{U}(\tilde{\bm{\theta}})\tilde{\bm{\Psi}}\tilde{\bm{\beta}} \|
\end{equation}
measuring the similarity between object and interference signals when projected onto the object signal base $\tilde{\bm{\Phi}}$, with equality reached around $-20<\mathrm{SIR}<-15\mathrm{dB}$. We hypothesize that such a condition has theoretical significance for the signal separation problem, but further rigorous analysis is necessary. The algorithm again tends to the \ac{crlb} at lower \ac{sir}. From a practical point of view, it is certainly most important in a radar application to achieve some guaranteed performance at critical low-\ac{sir} and low-\ac{snr} scenarios.

\subsection{Robustness to Interference Model Error}
\label{subsec:robustness_model_error}

\begin{figure}[ht]
\begin{center}
\scalebox{1}{\includegraphics{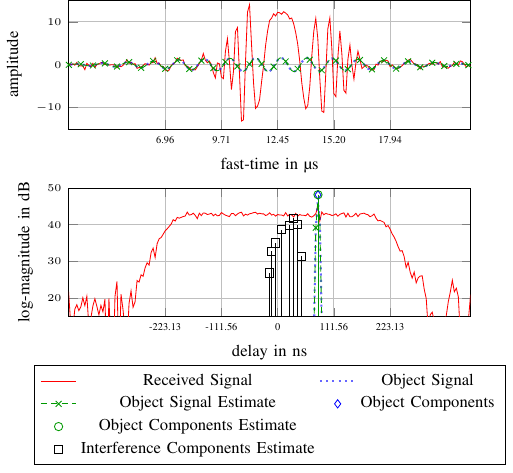}}
\caption{Example signal and estimation results at $\mathrm{SNR} = 30\mathrm{dB}$ and $\mathrm{SIR} = -15\mathrm{dB}$ with model error. The included signals are the same as for Fig.~\ref{fig:sig_estEx}. To note in particular is that the estimate for the interference channel is biased, meaning that it estimates more than a single component. This compensates for the model error, resulting in a more robust estimate of the object component.}
\label{fig:sig_estEx_aafError}
\end{center}
\end{figure}

Any model-based algorithm breaks down if the inference model is not sufficiently accurate to reality. In our application, it cannot for example be assumed that we have perfect knowledge of the \ac{aaf} transfer function $G(f)$ for specific radar hardware. Hence, we investigate the behavior of the algorithm when assuming, as previously, the raised cosine envelope for inference but in fact generating the tested interference signal using convolution with a Butterworth lowpass filter with equivalent filter parameters. This constitutes a small, but not negligible error in the filter's assumed transfer function. An example result is shown in Fig.~\ref{fig:sig_estEx_aafError}.

Fig.~\ref{fig:CRLB_group}c shows the same analysis as for Fig.~\ref{fig:CRLB_group}b, but with the introduced model error. As expected, an increasing error in the object channel parameter estimate becomes visible below a certain $\mathrm{SIR}$. Nevertheless, the algorithm does not abruptly break down and behaves robustly, reaching the \ac{crlb} for $\mathrm{SIR} \geq -30\mathrm{dB}$. The reason for this adaptability is illustrated in Fig.~\ref{fig:sig_estEx_aafError}. Note in Fig.~\ref{fig:sig_estEx_aafError} that several line spectral components are estimated for the interference channel, even though the signal is simulated with line-of-sight interference only. These additional components compensate for the \ac{aaf} model error.

\subsection{Statistical Performance Comparison}
\label{subsec:stat_comp}

\begin{figure*}[ht]
\begin{center}
\scalebox{1}{\includegraphics{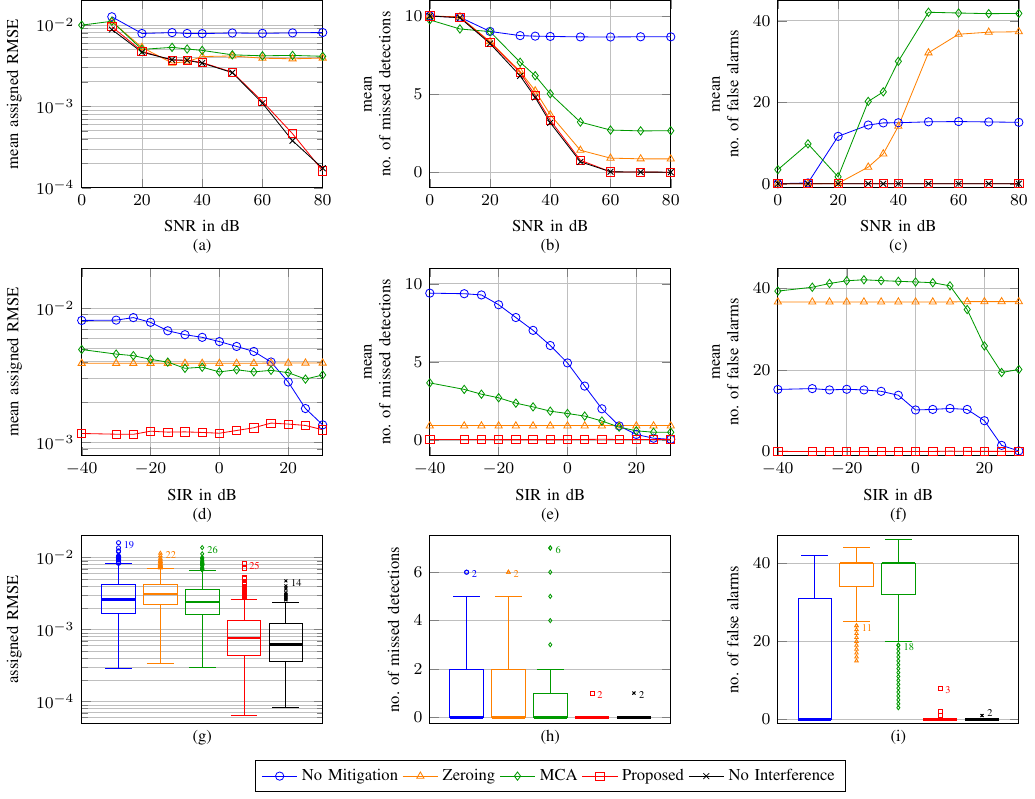}}
\caption{Statistical results of multi-object simulation scenario. (a)-(c) Sample mean over varying \ac{snr} at a fixed \ac{sir} of $-20\mathrm{dB}$. (d)-(f) Sample mean over varying \ac{sir} at a fixed \ac{snr} of $60\mathrm{dB}$. (g)-(i) Distribution of results at an \ac{snr} of $60\mathrm{dB}$ and \ac{sir} of $15\mathrm{dB}$ for each compared method, depicted in the form of box plots. Markers indicate outlier values, and the total number of outlier samples is additionally annotated.}
\label{fig:comp_group}
\end{center}
\end{figure*}

\begin{figure}[ht]
\begin{center}
\scalebox{1}{\includegraphics{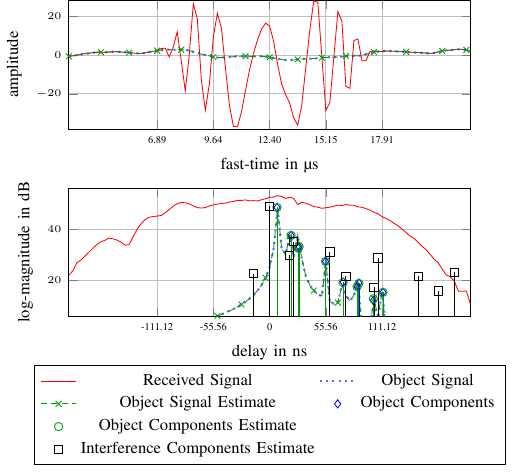}}
\caption{Example signals and estimates of a single interfered ramp at $\mathrm{SNR} = 60\mathrm{dB}$ and $\mathrm{SIR} = -15\mathrm{dB}$. The included signals are the same as for Fig.~\ref{fig:sig_estEx}.}
\label{fig:sig_estEx_comp}
\end{center}
\end{figure}

We study the performance of the proposed algorithm in broader, more complex scenarios. Table~\ref{tab:params} shows the relevant simulation parameters in the column denoted Simulation II. The radar system receives $N=128$ fast-time samples from $P=16$ ramps each, with the task to estimate objects' range and velocity parameters. The interferer transmit ramp parameters are chosen randomly for each realization from three different sets representing three scenarios. The three sets represent increasing interference burst durations while decreasing the number of interfered ramps, which arises naturally due to the form of \ac{fmcw} transmit signals.

Both object and interference channels consist of $\tilde{L}=\tilde{K}=10$ line spectral components. Dispersion parameters are chosen randomly on a uniform distribution, while component magnitudes are set proportionately according to the radar equation~\cite{richards_fundamentals_2014} with some random variation to account for the radar cross section of different physical objects. Note also the different exponents in the radar equation for victim and interference channels~\cite{schipper_estimation_2013}. For the interference, the first channel component is fixed to a delay of zero and relative magnitude of one. This is the direct path interference between sensors, which has been observed to dominate in power~\cite{schipper_simulation_2012}.

As these are multi-target scenarios, performance metrics appropriate for multi-target detection are selected. The \ac{gospa} metric~\cite{rahmathullah_generalized_2017} is suited to such cases. Its evaluation starts with an assignment of the estimated components to the true ones based on an upper-bounded estimation error metric. \ac{gospa} can be split into a so-called \textit{localization} and a \textit{cardinality} error term. The former is the sum of estimation errors for the assigned components, while the latter penalizes non-assigned detections. The cardinality error can be further split into two terms proportional to the number of misdetections and false alarms, respectively. We use the assignment method of \ac{gospa} and analyze variants of these three error terms, leading to an insightful set of performance metrics. For the localization error, we change the sum of errors of~\cite{rahmathullah_generalized_2017} to the mean of Euclidean distances. This yields a metric comparable to the single-component \ac{rmse}, here termed the \textit{mean assigned} \ac{rmse}. For the two cardinality error terms, we evaluate the absolute number of misdetections and false alarms directly.

Our proposed algorithm is compared with two other methods for interference mitigation found in literature. Note that these are \textit{preprocessing} methods to be employed prior to a separate object detection step in the automotive radar signal processing chain. In order to ensure a fair comparison, the ``object only'' algorithm of Section~\ref{subsec:fund_anal} is applied after preprocessing, as well as in the \textit{No Mitigation} and \textit{No Interference} test cases. The two methods, both introduced in Section~\ref{sec:introduction}, are:
\begin{itemize}
\item \textit{Zeroing} - zeroing relies on the explicit prior detection of interfered samples, which is assumed perfect here. Comparison of different algorithms for the detection of interfered samples is out of the scope of this work. 
\item \textit{\Acf{mca}} - using the concept and model of~\cite{uysal_synchronous_2018}. Our implementation follows that of~\cite[Algorithm 30]{starck_sparse_2010}.
\end{itemize}

Figs.~\ref{fig:comp_group}a-c depict the mean of the resulting metrics computed from 500 realizations over a varying \ac{snr}. Interference leads to a lower limit on the achievable estimation error and the number of misdetections, and causes false alarms. With preprocessing, the estimation error and number of misdetections improve, but the performance is still limited and improvements come at the cost of false alarms. This behavior can be explained by considering the effects of preprocessing on the form of the resulting signal. Zeroing merely exploits the time-limited nature of the interference, and hence leads to a signal with missing clusters of samples. \ac{mca} considers a signal separation problem, but does not employ a parametric model for the interference bursts and only applies single-ramp on-grid \ac{lse} for the object signal. Preprocessing therefore typically causes a distortion of the object signal that is a model error for the subsequent detector, primarily leading to false alarms. Our proposed method mitigates this behavior and reaches performance very close to the \textit{No Interference} case. The estimation error and the number of misdetections decrease with increasing \ac{snr}, and the number of false alarms is constant according to the set detection threshold.

Figs.~\ref{fig:comp_group}d-f show mean results at a high \ac{snr}, over \ac{sir} levels ranging from highly interfered to weakly interfered object signals. Ideally, perfect interference cancellation leads to values identical to the \textit{No Interference} case at $\mathrm{SNR}=60\mathrm{dB}$. The proposed method approximately achieves this up until $\mathrm{SIR} \approx 0\mathrm{dB}$. Its performance then slightly decreases with \textit{decreasing} interference power up until $\mathrm{SIR} \approx 15\mathrm{dB}$, after which it slowly recovers. This is the same effect as discussed in Section~\ref{subsec:fund_anal}. Note that unlike for the simulation scenario of Fig.~\ref{fig:CRLB_group}, in the scenario of Fig.~\ref{fig:comp_group} the condition of~\eqref{eq:coherence} is met roughly at $5\mathrm{dB}<\mathrm{SIR}<10\mathrm{dB}$. This strengthens our hypothesis on the significance of this condition. For the other evaluated methods, we see a bounded improvement in \ac{rmse} and misdetection rate at the cost of further false alarms. Naturally, perfect zeroing is independent of interference power while \ac{mca} performance is generally improved with increasing \ac{sir}.

Figs.~\ref{fig:comp_group}g-i illustrate the distributions of the results at $\mathrm{SNR} = 60\mathrm{dB}$ and $\mathrm{SIR} = -15\mathrm{dB}$ using box plots, in order to evaluate the \textit{robustness} of the tested methods. In particular for the proposed method, this further gauges the risk of the algorithm converging to a local optimum of the signal separation problem, which could significantly decrease object detection performance. We indeed observe some \textit{outliers} with higher \ac{rmse} values and an increased number of false alarms. However, the number of significant outliers, particularly when noting that the comparison is to be made to the ideal \textit{No Interference} test case, is very small.

We also include in Fig.~\ref{fig:sig_estEx_comp} an example plot analogous to Figs.~\ref{fig:sig_estEx} and~\ref{fig:sig_estEx_aafError}. We typically observe $\hat{K} > \tilde{K}$, i.e., the amount of interference channel line spectral components is overestimated in our interference estimation subroutine. This necessarily occurs when there are several relatively closely-spaced components, as we employ on-grid estimation on a small grid $(K=2N)$ with a relatively low detection threshold $T_\beta$. Nevertheless, it can be seen that the object signal is successfully estimated.

In summary, our algorithm statistically outperforms the investigated preprocessing methods across the whole investigated \ac{snr}/\ac{sir} space and only slightly diverges from optimal interference cancellation in a less critical high-\ac{sir} region of this exploration space.

\subsection{Measurement Examples}\label{subsec:meas_ex}

\begin{figure*}[ht]
\begin{center}
\scalebox{1}{\includegraphics{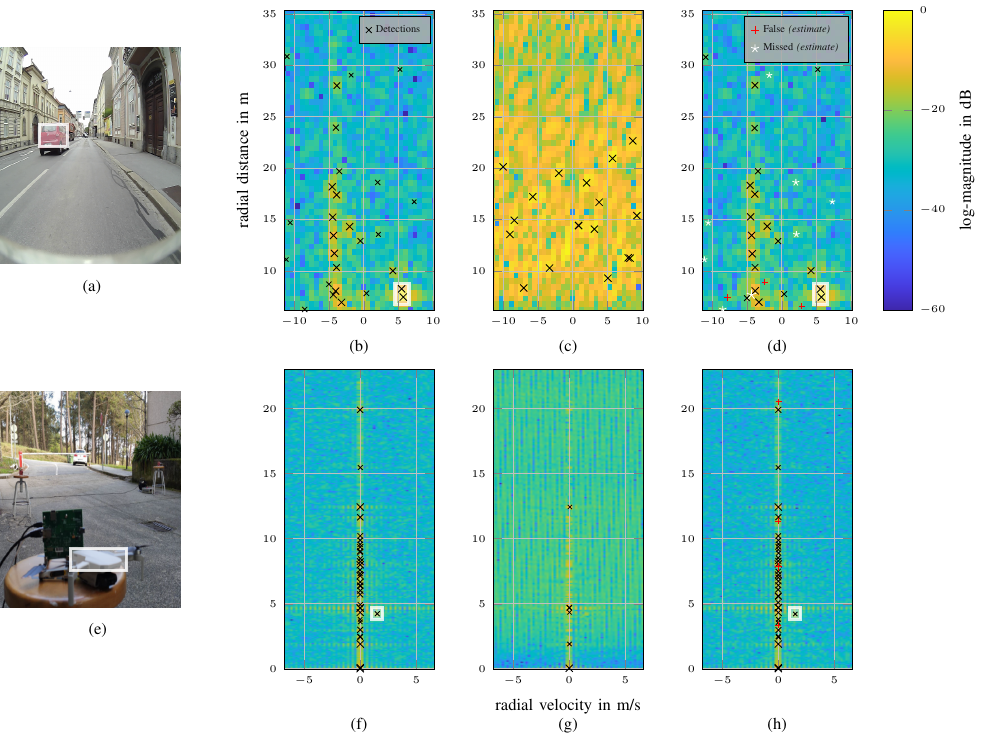}}
\caption{Two examples of the proposed algorithm applied to real automotive radar measurement data. The first row is from the authors' own combined object and interference measurements, while the second row is from a publicly available test dataset. White shaded rectangles indicate an object of interest that becomes undetectable due to interference. Detections are indicated by markers, where marker size represents component \ac{snr}. (a)/(e) Photo of the measurement scene. (b)-(c)/(f)-(g) Range-Doppler spectrum of the measured signal without/with interference and corresponding object detections. (d)/(h) Range-Doppler spectrum of the interference-cancelled signal and corresponding object detections as a result of the proposed algorithm. \textit{Estimated} false alarms and missed detections, when taking the results for the non-interfered measurement as ground truth, are also indicated.}
\label{fig:trafficMeas_RDmaps}
\end{center}
\end{figure*}

\begin{figure}[ht]
\begin{center}
\scalebox{1}{\includegraphics{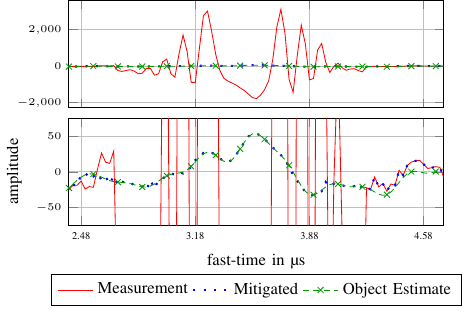}}
\caption{Qualitative example of the processed signal at a single interfered ramp and object signal estimation results. It shows a ramp of the measurement $\bm{r}$, of the mitigated signal being the object residual $\hat{\bm{r}}_{\alpha}$, and of the object signal estimate $\hat{\bm{r}}_{\mathrm{O}} = \hat{\bm{\Phi}}\hat{\bm{\alpha}}$. The top and bottom graphs are the same result with different y-axis limits, the latter showing the much lower amplitude object signal.}
\label{fig:trafficMeas_exSig}
\end{center}
\end{figure}

In order to illustrate the practical application of the proposed method, we apply it to radar measurement data from two different datasets. We select a single measurement each and showcase the results qualitatively; note that no ground truth of the channel parameters are available. 

The first example was obtained as part of a measurement campaign in real traffic scenarios, details of which can be found in previous works~\cite{toth_analysis_2021,rock_deep_2020}. Measurement parameters are summarized in Table~\ref{tab:params} (Obj. Measurement). These measurements do not inherently contain any interference. In prior work, simulated interference was added to the measured signals. In this work however, measured interference is added, as described in Appendix~\ref{sec:appendix:meas_data}, with parameters listed in Table~\ref{tab:params} (Int. Measurement). The interference was obtained from a separate measurement, due to the difficulty of generating and measuring interference separately in the same inner city traffic environment. Nevertheless, the applied scheme enables a qualitative comparison of our mitigation result to an interference-free one, with both object and interference components obtained using actual radar sensors.

Fig.~\ref{fig:trafficMeas_RDmaps}a shows a photo of the selected Obj. Measurement scene. Particularly, visible on the photo is the car in front and to the left of the ego-radar, as well as the facade of a row of buildings to the right. The interference-free spectrum in Fig.~\ref{fig:trafficMeas_RDmaps}b hence contains a few close object components with a positive relative velocity, i.e., the car moving away from the sensor. The buildings lead to a large number of components distributed over distances, with negative relative velocities with respect to the measurement vehicle. Other detected components are reflections not easily associated with the camera image or false alarms due to noise. When strong interference is introduced, the range-Doppler spectrum becomes dominated by it, see Fig.~\ref{fig:trafficMeas_RDmaps}c. Directly applying object detection onto this signal leads to no discernibly useful result. Finally, the spectrum of the measurement when subtracting the interference estimated by the proposed processing algorithm, and the estimated object components are shown in Fig.~\ref{fig:trafficMeas_RDmaps}d. The resulting spectrum is visually very close to the interference-free one and the detected components correspond well to the significant parts of the described scene. A number of missed detections and false alarms are observed, but mostly of comparatively small magnitudes. These are likely in large part a result of an interference model error due to unknown measurement hardware characteristics, further discussed in Appendix~\ref{sec:appendix:meas_data}. We also qualitatively illustrate the fast-time signal of a single ramp in Fig.~\ref{fig:trafficMeas_exSig} as an example. The top graph shows a large interference burst not too dissimilar to, but distinct from the simulated ones of Figs.~\ref{fig:sig_estEx} and~\ref{fig:sig_estEx_aafError}. The bottom plot provides a close-up view of the signal after subtraction of the interference estimate as well as shows the object signal estimate.

The second example measurement originates from~\cite{lopez_rawdata_2023, lopez-valcarcel_mti-like_2023}. Measurement parameters and further details are found in the cited references. The measurement hardware and scenario are distinct from the previously presented measurement, in particular with respect to the interference. Here, the measurements inherently contain interference from two different sensors, so that no additional processing as described in Appendix~\ref{sec:appendix:meas_data} is necessary. The scenario is mostly static with significant reflections up to a distance of $20\mathrm{m}$, as well as a small drone flying around. Figs.~\ref{fig:trafficMeas_RDmaps}e-h show the results for this example in an analogous manner to Figs.~\ref{fig:trafficMeas_RDmaps}a-d. No exact ground truth is available, but the ``clean'' reference of Fig.\ref{fig:trafficMeas_RDmaps}f is taken from the previous measurement frame, during which no interferer crossed the victim radar's \ac{if} band. Similarly to the previous example, we can see that the mitigated spectrum as well as the detected components closely match the results for the interference-free measurement. We also observe a few false alarms, although no missed detections in contrast to the previous example measurement. This may largely be the consequence of the interference being significantly weaker in this scenario.

These very promising results are nevertheless only preliminary, and a larger scale analysis with better known hardware and measurement environment is a necessary part of future work.

\section{Conclusion}
\label{sec:conclusion}

This paper presents an algorithm to mitigate the mutual interference of automotive \ac{fmcw} radars. We fully consider the underlying signal model and systematically design a model-based inference algorithm. Our description of the interference considers both sensor-based effects and the multipath propagation environment. We propose to infer the delay-Doppler object channel and ramp-wise delay-only interference channels. The result is a superposition of \ac{lse} problems, with the non-coherent interference chirp envelopes modeled by additional unknown parameters. We extend the state-of-the-art sparse probabilistic inference approach to such a superposition within the variational \ac{em} framework. Our proposed inference model leads to an iterative algorithm consisting of subsequent object and interference estimation subroutines. This results in robust object detection performance that is often comparable to the interference-free case. Conventional interference mitigation preprocessing is shown to often lead to increasing false alarms in the object detection step, which the proposed algorithm is not susceptible to, in comparison. It is also shown that considering the estimation of interference as an \ac{lse} problem can offset minor model errors.

Specific choices in the proposed algorithm design, initialization and scheduling are essential for its performance. This includes employing grid-less estimation for the object channel, yielding super-resolution accuracy, while using an on-grid algorithm for the interference channels. Parameter optimization and inference of the respective channel weights are derived from a joint update, which decreases the probability of converging to a local optimum of the signal separation problem.

The proposed method has some limitations due to assumptions made in its derivation, as elaborated in the paper. Most prominently, at most a single interference chirp is assumed per ramp, and interference chirps are present over the full passband of the receiver's \ac{aaf}. While ways to adapt the algorithm are briefly discussed, more work is needed to verify their practical viability in different scenarios. Furthermore, we posit that signal separability is fundamentally limited by how distinct the involved signal bases are. Investigating such limits is also an interesting prospect for further work. Finally, note that although here presented for a specific application, the proposed algorithm is likely applicable to other signal separation problems of this kind.

\FloatBarrier
\appendices
\section{Derivation of Signal Model}
\label{sec:appendix:interference_model}

\subsection{Object Signal}

To derive~\eqref{eq:IF_signal_victim}, we start with the transmit signal of~\eqref{eq:victim_transmit_signal}. It propagates through the environment and reflects off of (moving) \textit{objects}, i.e., a weighted sum of delayed transmit signals reach the receiver. The radar echo as a function of time $t$ can be derived for ideal point reflections starting with $y_{\mathrm{O}}(t) = \sum_{l=0}^{\tilde{L}-1} \tilde{\alpha}_l x(t-\tilde{\tau}_l(t))$, where $\tilde{\tau}_l(t) \approx 2(d_l + v_l t)/\mathrm{c}_0$ with $d_l$ and $v_l$ being radial distance and velocity~\cite{stove_linear_1992}, and $\mathrm{c}_0$ denotes the speed of light. In our derivations, we use an analogous formulation based on channel modeling~\cite{hlawatsch_wireless_2011}. I.e., the transmit signal propagates through the radar channel which can be described by the operation
\vspace*{-1mm}
\begin{align}
y_{\mathrm{O}}(t) &= \int_{\nu} \int_{\tau} h_{\mathrm{O}}(\tau,\nu) x(t-\tau) \exp{(\mathrm{j}2\pi \nu t)} \mathrm{d}\tau \mathrm{d}\nu \nonumber\\
&= \sum_l \tilde{\alpha}_l x(t-\tilde{\tau}_l) \exp{(\mathrm{j}2\pi \tilde{\nu}_l t)}
\label{eq:int_model_obj_channel}\\[-7mm]\nn
\end{align}
with
\vspace*{-1mm}
\begin{align}
h_{\mathrm{O}}(\tau,\nu) = \sum^{\tilde{L}-1}_{l=0} \tilde{\alpha}_l\delta(\tau-\tilde{\tau}_l)\delta(\nu-\tilde{\nu}_l) \label{eq:int_model_obj_channel2l}\\[-7mm]\nn
\end{align}
where the channel is described by its \textit{spreading function} $h_{\mathrm{O}}(\tau,\nu)$ in delay $\tau$ and Doppler frequency $\nu$. The model assumes negligible \textit{coupling} between delay and Doppler dispersion domains. Expanding on~\eqref{eq:int_model_obj_channel}, we have
\vspace*{-1mm}
\begin{align}
{\hspace{-0.2cm}} y_{\mathrm{O}}(t) &= A \sum^{\tilde{L}-1}_{l=0} \sum^{P-1}_{p=0} \bar{x}(t-T_{\mathrm{p}} p-\tilde{\tau}_l; T_\mathrm{sw}) \ist \tilde{\alpha}_l \ist \exp{(-\mathrm{j}2\pi \tilde{\nu}_l t)} \nonumber\\ & \times \exp{(\mathrm{j}(2\pi f_0 (t-T_{\mathrm{p}} p-\tilde{\tau}_l) + \pi k (t-T_{\mathrm{p}} p-\tilde{\tau}_l)^2))} \\[-7mm]\nn
\end{align}
where we will assume $A = 1$ for simplicity, without loss of generality.

To simplify this, we first note that in a continuous wave radar system, the maximum effective delay $\mathrm{max}(\tilde{\tau}_l)$ of the channel is \emph{much} shorter than the duration of a transmit ramp $T_{\mathrm{sw}}$. We can hence approximately neglect the spreading of the ramp envelope itself. I.e., we swap the order of summations and apply $\bar{x}(t-T_{\mathrm{p}} p-\tilde{\tau}_l; T_\mathrm{sw}) \approx \bar{x}(t-T_{\mathrm{p}} p; T_\mathrm{sw})$. Furthermore, the \textit{stop-and-go} approximation $\tilde{\nu}_l t \approx \tilde{\nu}_l T_\mathrm{p}p$ as commonly known in automotive radar~\cite{richards_fundamentals_2014} is applied. We obtain
\vspace*{-1mm}
\begin{align}
{\hspace{-0.1cm}} y_{\mathrm{O}}(t) &\approx \sum^{P-1}_{p=0} \bar{x}(t-T_{\mathrm{p}} p; T_\mathrm{sw})\ist \sum^{\tilde{L}-1}_{l=0} \ist \tilde{\alpha}_l \ist \exp{(-\mathrm{j}2\pi \tilde{\nu}_l T_\mathrm{p}p)}  \nn\\ &\times \exp{(\mathrm{j}(2\pi f_0 (t-T_{\mathrm{p}} p-\tilde{\tau}_l) + \pi k (t-T_{\mathrm{p}} p-\tilde{\tau}_l)^2))} \,. \\[-7mm]\nn
\end{align}
Multiplying out and rearranging the terms in the chirp exponential straightforwardly leads to
\vspace*{-1mm}
\begin{align}
y_{\mathrm{O}}(t) &= \sum^{P-1}_{p=0} \bar{x}(t-T_{\mathrm{p}} p; T_\mathrm{sw}) \nn\\ & \, \times \exp(\mathrm{j}(2\pi f_0 (t-T_{\mathrm{p}} p) +\pi k (t-T_{\mathrm{p}} p)^2))) \nn\\ & \, \times \sum^{\tilde{L}-1}_{l=0} \tilde{\alpha}_l \exp{(\mathrm{j}(-2\pi (f_0 + k (t-T_{\mathrm{p}} p)) \tilde{\tau}_l )} \nn\\ & \quad \times \exp{(\mathrm{j}(\pi  k \tilde{\tau}^2_l ))} \exp{(-\mathrm{j}2\pi \tilde{\nu}_l T_\mathrm{p}p)} \,.
\label{eq:appendix_sObjChannel} \\[-7mm]\nn
\end{align}
Reasoning again based on the short duration of the channel, we neglect the term $k \tilde{\tau}^2_l \approx 0 \forall l$. The rest of the expression in the last brackets is then the Fourier transform of the channel spreading function. Defining this as the object channel transfer function $H_{\mathrm{O}}(f,t)$, we write~\footnote{Somewhat uncommonly for automotive radar literature, in our formulation the fast-time signal measures the frequency domain of the channel whose \emph{inverse} Fourier spectrum represents the delay dispersion. Note that delay and frequency are always dual domains of the radio channel, whereas delay being \textit{proportional} to a \textit{beat frequency} is a specific ``quirk'' of \ac{fmcw} radar.}
\vspace*{-1mm}
\begin{align}
y_{\mathrm{O}}(t) &\approx \sum^{P-1}_{p=0} \bar{x}(t-T_{\mathrm{p}} p; T_\mathrm{sw}) \exp(\mathrm{j}(2\pi f_0 (t-T_{\mathrm{p}} p) \nn\\ &  + \pi k (t-T_{\mathrm{p}} p)^2)) H_{\mathrm{O}}(f_0 + k (t-T_{\mathrm{p}} p), T_{\mathrm{p}} p) \,. \\[-7mm]\nn
\end{align}

Next, demodulation \textit{(demixing)} according to \textit{stretch processing}~\cite{caputi_stretch_1971} is applied. Demixing is coherent for the object signal, so the transmit ramp envelope $\bar{x}(t)$ over the ramp duration stays constant and the chirp term vanishes. I.e.,
\vspace*{-1mm}
\begin{align}
{\hspace{-0.2cm}}y^{\prime}_{\mathrm{O}}(t) &= \biggl( x^{\ast}(t) \sum^{P-1}_{p=0} \bar{x}(t-T_{\mathrm{p}} p; T_\mathrm{sw}) \exp(\mathrm{j}(2\pi f_0 (t-T_{\mathrm{p}} p) \nn\\[-2mm] 
& \quad + \pi k (t-T_{\mathrm{p}} p)^2)) H_{\mathrm{O}}(f_0 + k (t-T_{\mathrm{p}} p), T_{\mathrm{p}} p) \biggr) \nn\\[-2mm]
&= \sum^{P-1}_{p=0} \bar{x}(t-T_{\mathrm{p}} p; T_\mathrm{sw}) H_{\mathrm{O}}(f_0 + k (t-T_{\mathrm{p}} p), T_{\mathrm{p}} p) \,. \\[-7mm]\nn
\end{align}

Finally, we assume that the object channel contains no significant amount of energy outside of the constant passband of the \ac{aaf}, hence the convolution with $g(t)$ can be neglected. The signal is projected onto the fast-time $t^{\prime} \in (0,T_\mathrm{sw}) = t-T_\mathrm{p} p \, \forall p$ and slow-time $p$ so that the explicit sum over $p$ vanishes as well. This leads to the result in~\eqref{eq:IF_signal_victim}.\footnote{This is a standard result that is well established. Nevertheless, its detailed derivation within our channel modeling-based formulation in this work is warranted to set the derivation of the interference signal into the proper context.}

\subsection{Interference Signal}

We derive the received interference signal from a single interferer, $r_{\mathrm{I},i}(t^{\prime},p)$ of~\eqref{eq:IF_signal_interference}, starting from the interferer transmit signal $x_{\mathrm{I},i}(t)$ of~\eqref{eq:transmit_sig_int}. Note that for notational simplicity, we drop the explicit interferer subscript $i$ in the sequel. First, propagation through a multipath channel leads to the expression
\vspace*{-1mm}
\begin{align}
{\hspace{-0.2cm}} y_{\mathrm{I}}(t) &\approx x_{\mathrm{I}}(t) \sum^{\tilde{K}-1}_k \tilde{\beta}_k \exp(-\mathrm{j}2\pi (f_{\mathrm{I}} + k_{\mathrm{I}}(t-\bar{T}-T_{\mathrm{pI}}p_{\mathrm{I}})) \tilde{\tau}_{\mathrm{I},k}) \nn\\[-2mm] &{\hspace{2.5cm}} \times \exp{(-\mathrm{j}2\pi T_{\mathrm{pI}}p_{\mathrm{I}} \tilde{\nu}_{\mathrm{I},k} )} \nn\\[1mm] &= x_{\mathrm{I}}(t) H_\mathrm{I}(f_\mathrm{I}+k_\mathrm{I}(t-\bar{T}-T_{\mathrm{pI}}p_\mathrm{I}),T_{\mathrm{pI}}p_\mathrm{I})
\label{eq:int_through_channel} \\[-7mm]\nn
\end{align}
analogously as for the object signal $y_{\mathrm{O}}(t)$. However, this signal is non-coherently demodulated by multiplication with $x^{\ast}(t)$. To elaborate, we explicitly write out the mixing, i.e.,
\begin{align}
y^{\prime}_\mathrm{I}(t) &= \bar{A} \sum^{P-1}_{p=0} \bar{x}(t-T_{\mathrm{p}} p; T_\mathrm{sw}) \nn\\ & \quad \times \exp(-\mathrm{j}(2\pi f_0 (t-T_{\mathrm{p}} p) + \pi k (t-T_{\mathrm{p}} p)^2)) \nn\\ & \quad \times \biggl( \sum^{P_\mathrm{I}-1}_{p_\mathrm{I}=0} \bar{x}(t-\bar{T}-T_{\mathrm{pI}} p_\mathrm{I}; T_\mathrm{swI}) \nn\\ & \quad\quad \times \exp(\mathrm{j}(2\pi f_\mathrm{I} (t-\bar{T}-T_{\mathrm{pI}} p_\mathrm{I}) \nn\\ & \quad\quad\quad + \pi k_{\mathrm{I}}(t-\bar{T}-T_{\mathrm{pI}} p_\mathrm{I})^2)) \nn\\ & \quad\quad \times H_\mathrm{I}(f_\mathrm{I}+k_\mathrm{I}(t-\bar{T}-T_{\mathrm{pI}}p_\mathrm{I}),T_{\mathrm{pI}}p_\mathrm{I}) \biggr)
\end{align}
setting $\bar{A} = A_{\mathrm{I}}A^{\ast} = 1$ without loss of generality in the sequel.

For convenience, we rewrite this in the fast- and slow-time domain description by substituting $t = t^{\prime} + T_\mathrm{p} p$, leading to
\begin{align}
y^{\prime}_\mathrm{I}(t^{\prime},p) &= \exp(-\mathrm{j}(2\pi f_0 t^{\prime} + \pi k {t^{\prime}}^2) ) \nn\\ 
& \times \sum^{P_\mathrm{I}-1}_{p_\mathrm{I}=0} \bar{x}(t^{\prime} +T_\mathrm{p} p -\bar{T} -T_{\mathrm{pI}} p_\mathrm{I}; T_\mathrm{swI}) \nn\\ 
& \times \exp(\mathrm{j}(2\pi f_\mathrm{I} (t^{\prime} + T_\mathrm{p} p-\bar{T}-T_{\mathrm{pI}} p_\mathrm{I}) \nn\\ 
&{\hspace{0.5cm}} + \pi k_{\mathrm{I}}(t^{\prime} +T_\mathrm{p} p-\bar{T}-T_{\mathrm{pI}} p_\mathrm{I})^2)) \nn\\ 
& \times H_\mathrm{I}(f_\mathrm{I}+k_\mathrm{I}(t^{\prime} +T_\mathrm{p} p -\bar{T}-T_{\mathrm{pI}}p_\mathrm{I}),T_{\mathrm{pI}}p_\mathrm{I})
\end{align}
noting again that the sum over $p$ vanishes.

Carrying out the multiplication of the victim and interferer transmit chirps and rearranging the resulting phase terms with respect to $t^{\prime}$-dependency yields
\begin{align}
{\hspace{-0.25cm}} y^{\prime}_\mathrm{I}(t^{\prime},p) &= \sum^{P_\mathrm{I}-1}_{p_\mathrm{I}=0} \bar{x}(t^{\prime} +T_\mathrm{p} p -\bar{T} -T_{\mathrm{pI}} p_\mathrm{I}; T_\mathrm{swI}) \nn\\ 
&\times \exp(\mathrm{j}(2\pi (f_\mathrm{I} + k_\mathrm{I}(T_\mathrm{p} p - \bar{T} - T_\mathrm{pI} p_\mathrm{I}) - f_0 ) t^{\prime} \nn\\ 
&{\hspace{0.5cm}} + \pi (k_\mathrm{I} - k) {t^{\prime}}^2 )) \nn\\ 
&\times \exp(\mathrm{j}(2\pi f_\mathrm{I} (T_\mathrm{p} p-\bar{T}-T_{\mathrm{pI}} p_\mathrm{I}) \nn\\ 
&{\hspace{0.5cm}} + \pi k_{\mathrm{I}}(T_\mathrm{p} p-\bar{T}-T_{\mathrm{pI}} p_\mathrm{I})^2)) \nn\\
& \times H_\mathrm{I}(f_\mathrm{I}+k_\mathrm{I}(t^{\prime} +T_\mathrm{p} p -\bar{T}-T_{\mathrm{pI}}p_\mathrm{I}),T_{\mathrm{pI}}p_\mathrm{I})\,.
\label{eq:r_mix_detail}
\end{align}

Due to non-coherent demodulation, the effect of the \ac{aaf} cannot be neglected for the interference. We can approximate the \ac{aaf} for every slow-time index $p$ separately as the convolution with an impulse response $g(t^{\prime})$. We define $\bar{T}_0 = T_\mathrm{p} p -\bar{T} -T_{\mathrm{pI}} p_\mathrm{I}$ to de-clutter notation, as well as the non-coherent demixed chirp
\begin{align}
\bar{u}_{\mathrm{pI}}(t^{\prime},p) &= \exp(\mathrm{j}(2\pi \Delta f_0 t^{\prime} + \pi \Delta \tilde{k} {t^{\prime}}^2 )) \nn\\ 
&\hspace{5mm} \times \exp\bigl(\mathrm{j}(2\pi f_\mathrm{I} \bar{T}_0 + \pi k_{\mathrm{I}} \bar{T}_0^2) \bigr)
\label{eq:noncoh_int_chirp}
\end{align}
where $\Delta f_0 = f_\mathrm{I}-f_0$ and $\Delta \tilde{k} = k_\mathrm{I} - k$. This leads to the convolution integral expression
\begin{align}
r_\mathrm{I}(t^\prime, p) &= \sum^{P_\mathrm{I}-1}_{p_\mathrm{I}=0} \bar{u}_{\mathrm{pI}}(t^{\prime},p) \biggl( \int_\varsigma \bar{x}(t^{\prime}-\varsigma +\bar{T}_0; T_\mathrm{swI}) \nn\\ 
& \times \exp \bigl(-\mathrm{j} (2\pi (\Delta f_0 + k_\mathrm{I} \bar{T}_0) \varsigma + 2\pi \Delta \tilde{k} t^{\prime} \varsigma - \pi \Delta \tilde{k} \varsigma^2 ) \bigr) \nn\\ 
& \times H_\mathrm{I}(f_\mathrm{I}+k_\mathrm{I}( (t^{\prime}-\varsigma) +\bar{T}_0),T_{\mathrm{pI}}p_\mathrm{I}) g(\varsigma) \, \mathrm{d}\varsigma \biggr)\,.
\label{eq:noncoh_int_conv_full}
\end{align}

Incorporating~\eqref{eq:noncoh_int_conv_full} directly into the inference model is possible, but leads to significantly increased complexity. In order to derive a simplified form used in this work, certain further approximations are considered. We introduce here the first two out of four main assumptions employed for inference:
\begin{description}
\item[\textbf{Assumption 1}] \hspace{1.25cm} The envelope term $\bar{x}(t^{\prime}-\varsigma +\bar{T}_0; T_\mathrm{swI})$ can be neglected if it is assumed to be long enough in comparison to the filter impulse response. This assumption is correct in cases where the interferer's transmit ramp is active during the whole time its frequency course would be inside the victim receiver's \ac{if} bandwidth.
\item[\textbf{Assumption 2}] \hspace{1.25cm} We take the channel term $H_\mathrm{I}$ outside the convolution integral, neglecting its dependence on $\varsigma$. This is based on the reasoning that the chirp term $\bar{u}_{p\mathrm{I}}$ dominates the result of the convolution.
\end{description}
We then rearrange the remaining chirp terms and obtain
\begin{align}
\hspace*{-1mm}r_\mathrm{I}(t^\prime, p)\rmv &\approx\rmv\rmv  \sum^{P_\mathrm{I}-1}_{p_\mathrm{I}=0}\rmv\rmv\rmv \bar{u}_{\mathrm{pI}}(t^{\prime},p) H_\mathrm{I}(\bar{f}_\mathrm{I}+k_\mathrm{I} t^{\prime},T_{\mathrm{pI}}p_\mathrm{I}) \rmv\int_\varsigma \rmv\rmv\exp(\mathrm{j}(\pi \Delta \tilde{k} \varsigma^2 )) \nn\\ 
& \times \rmv g(\varsigma)\exp(\mathrm{-j}(2\pi (\Delta \tilde{f}_0 \varsigma + \Delta \tilde{k} t^{\prime} \varsigma ))) \, \mathrm{d}\varsigma
\end{align}
where we have further defined $\bar{f}_\mathrm{I} = f_\mathrm{I} + k_\mathrm{I} \bar{T}_0$ and $\Delta \tilde{f}_0 = \Delta f_0 + k_\mathrm{I} \bar{T}_0$.

The non-coherently demixed signal with which the \ac{aaf} impulse response is convolved is a chirp. Thus, similarly to the derivation for the multipath channel, we may describe the convolution result as approximately a multiplication with the \ac{aaf} frequency transfer function evaluated over the frequency course of this chirp. However, the term $\exp(\mathrm{j}(\pi \Delta \tilde{k} \varsigma^2 ))$ cannot in general be neglected as the length of the \ac{aaf} impulse response is long enough so that $\Delta \tilde{k} \varsigma^2 \not\approx 0$. Hence, we define the \textit{modified} \ac{aaf} impulse response as $\bar{g}(t^{\prime}) = g(t^{\prime}) \exp(\mathrm{j}(\pi \Delta \tilde{k} {t^{\prime}}^2 ))$ and its frequency transfer function $\bar{G}(f)$. In this way, the remaining expression is of an analogous form to~\eqref{eq:appendix_sObjChannel}, finally yielding~\eqref{eq:IF_signal_interference}.

\subsection{Inference Model}

While our derivations yield parametric models for all terms of~\eqref{eq:IF_signal_full}, the full model contains an unknown number of interferers. These cause interference bursts distributed over the received ramps depending on the victim and interferer transmit parameters as well as the propagation channel. To relax the resulting inference problem, we make the following further model assumptions in this work:
\begin{description}
\item[\textbf{Assumption 3}] \hspace{1.25cm} We limit the maximum amount of interference bursts at a certain slow-time index $p$ to one, i.e., the sum over $p_\mathrm{I}$ in~\eqref{eq:IF_signal_interference} vanishes. \\ If the number of bursts is known or estimated a-priori, our algorithm can be trivially adapted, although signal separation becomes increasingly ill-posed with a growing number of bursts. If it is unknown, it may be included in the inference model by considering a \textit{structured} dictionary~\cite{wipf_sparse_2011}. Algorithms for problems of this kind within the variational framework have been proposed in recent works~\cite{moderl_variational_2023, moderl_fast_2023}.
\item[\textbf{Assumption 4}] \hspace{1.25cm} We consider every interference burst over the slow-time independently. The channel term in~\eqref{eq:r_mix_detail} therefore becomes for every ramp $p$ a different delay-only line spectrum with frequency transfer function $H^{(p)}_{\mathrm{I}}(n)$. \\ For a single interferer, it is possible to consider the interference signal as the sequence of all received bursts propagating through a delay-Doppler channel. We have not investigated such an approach in this work, and note that more complex model assumptions can increase the risk of major model errors, as well as lead to increased algorithm complexity.
\end{description}

The above assumptions together mean that the sum $\sum^{M_{\mathrm{I}}-1}_{i=0} r_{\mathrm{I},i}(n,p)$ can be replaced by a single equivalent $r_{\mathrm{I}}(n,p)$ irrespective of $M_{\mathrm{I}}$. Furthermore, terms of this interference signal $r_{\mathrm{I}}(n,p)$ that do not explicitly depend on the fast-time index $n$ become constants for inference. These terms include the second complex phasor factor of~\eqref{eq:noncoh_int_chirp}, as well as certain terms of the transfer functions $H_\mathrm{I}(f)$ and $\bar{G}(f)$. These constants are simply inferred within the complex channel weights $\bm{\beta}^{(p)}$. Hence, the model is finally rewritten as~\eqref{eq:r_line}.

\section{Mean-field Variational EM Updates}
\label{sec:appendix:fast_updates}

For the structured mean-field proxy \ac{pdf}  $q = \prod_{\mathcal{Q}} q_i$, the \ac{elbo}~\eqref{eq:elbo} can be recast to express its dependence of any one factor $q_i$ as~\cite{bishop2009,tzikas_variational_2008}
\vspace*{-1mm}
\begin{align}
\mathcal{L}(q) &= \expect{\log p{(\bm{r},\bm{\alpha},\bm{\gamma}_{\alpha},\bm{\beta},\bm{\gamma}_{\beta},\lambda ; \bm{\zeta},\bm{\theta})}}{q} - \expect{\log q}{q} \nn\\ 
&= \expect{\expect{\log p{(\bm{r},\bm{\alpha},\bm{\gamma}_{\alpha},\bm{\beta},\bm{\gamma}_{\beta},\lambda ; \bm{\zeta},\bm{\theta})}}{\bar{q}_i}}{q_i} \nn\\ 
& \hspace{1cm} - \expect{\log q_i}{q_i} - E_i \nn\\
&= Z_i + \expect{\log \bar{p}_i}{q_i} - \expect{\log q_i}{q_i} - E_i \\[-7mm]\nn
\end{align}
\vspace*{-1mm}
with
\begin{align}
E_i &= \sum_{\mathcal{Q} \setminus q_i} \expect{\log q_j}{q_j} \\
\bar{p}_i &= {Z_i}^{-1} \exp{\left( \expect{\log p{(\bm{r},\bm{\alpha},\bm{\gamma}_{\alpha},\bm{\beta},\bm{\gamma}_{\beta},\lambda ; \bm{\zeta},\bm{\theta})}}{\bar{q}_i} \right)} \\
Z_i &= \int \exp{\left( \expect{\log p{(\bm{r},\bm{\alpha},\bm{\gamma}_{\alpha},\bm{\beta},\bm{\gamma}_{\beta},\lambda ; \bm{\zeta},\bm{\theta})}}{\bar{q}_i} \right)} \mathrm{d}\bm{z}_i \label{eq:Z_i} \\[-7mm]\nn
\end{align}
where $\bm{z}_i$ denotes the latent variables corresponding to the factor $q_i$. With all other latent variables $\bar{q}_i$ \textit{as well as parameters} fixed, the term $E_i$ and the normalization factor $Z_i$ are both independent of $q_i$. The remaining terms form a \textit{negative} Kullback-Leibler divergence (KLD), which is maximized for $q_i \propto \bar{p}_i$, yielding~\eqref{eq:mean_field_update}. When only $\bar{q}_i$ are fixed (i.e., parameters are to be optimized jointly with $q_i$), $E_i$ is  still a constant, but $Z_i$ is a function of the parameters. The properties of the negative KLD term remain, still yielding the solution $q_i \propto \bar{p}_i$ for the factor proxy \ac{pdf}. Hence, the \ac{elbo} is maximized when $Z_i$ is maximum, which directly leads to~\eqref{eq:joint_param_update} for parameter estimation.

We proceed to summarize the solution steps needed to obtain the update equations for our model. To clarify the mathematical derivations to follow, we write out the logarithms of factor distributions in the joint \ac{pdf}~\eqref{eq:jointpdf_general}, i.e.~\eqref{eq:likelihood_dist} to~\eqref{eq:lambda_dist}, explicitly so that
\vspace*{-1mm}
\begin{align}
\log p(\bm{r} \vert \bm{\alpha}, & \bm{\beta}, \lambda ; \bm{\zeta}, \bm{\theta}) \nn \\ &=  -PN\log\pi + PN\log\lambda \nonumber\\ & \,\quad -\lambda{(\bm{r}-\bm{\Phi\alpha}-\bm{U\Psi\beta})}^{\mathrm{H}}(\bm{r}-\bm{\Phi\alpha}-\bm{U\Psi\beta}) \label{eq:likelihood_dist_explicit}\\
\log p{(\bm{\alpha} \vert \bm{\gamma}_{\alpha})} &= \sum_l \Bigl( -\log\pi + \log\gamma_{\alpha,l} - {\vert \alpha_{l} \vert}^2\gamma_{\alpha,l} \Bigr) \label{eq:alpha_dist_explicit}\\
\log p{(\bm{\gamma}_{\alpha})} &= (a_{0}-1)\log\gamma_{\alpha,l} - b_0 \gamma_{\alpha,l} \nonumber\\ & \,\quad\quad + a_0 \log b_0 - \log(\Gamma(a_0)) \label{eq:gammaAlpha_dist_explicit}\\
\log p{(\bm{\beta} \vert \bm{\gamma}_{\beta})} &= \sum_p \sum_k \Bigl( -\log\pi + \log\gamma^{(p)}_{\beta,k} - {\vert \beta^{(p)}_k \vert}^2\gamma^{(p)}_{\beta,k} \Bigr) \label{eq:beta_dist_explicit}\\
p{(\bm{\gamma}_{\beta})} &= (c_0-1)\log\gamma^{(p)}_{\beta,k} - d_0\gamma^{(p)}_{\beta,k} \nonumber\\ & \,\quad\quad + c_0\log d_0 - \log(\Gamma(c_0)) \label{eq:gammaBeta_dist_explicit}\\
p(\lambda) &= (e_0-1)\log\lambda - f_0\lambda \nonumber\\ & \,\quad\quad + e_0\log f_0 - \log(\Gamma(e_0))\,. \label{eq:lambda_dist_explicit}
\end{align}
We refer back to Section~\ref{sec:signal-model} for the introduction of this probabilistic model and its components.

\subsection{Updates of Channel Weights}

Using~\eqref{eq:mean_field_update} to solve for the object channel weights $q_\alpha$, first note that the expectation of~\eqref{eq:jointpdf_general} (i.e., the sum of~\eqref{eq:likelihood_dist_explicit} to~\eqref{eq:lambda_dist_explicit}) with respect to $q_\lambda$ and the $q_\gamma$'s is obtained simply by applying $\lambda \rightarrow \hat{\lambda}$ and $\gamma \rightarrow \hat{\gamma}$. Neglecting constant terms independent of $\bm{\alpha}$, we are left with
\vspace*{-1mm}
\begin{multline}\label{eq:log_qalpha_1}
\log q_\alpha \propto^{\mathrm{e}} -\hat{\lambda}\expect{(\bm{r}-\hat{\bm{\Phi}}\bm{\alpha}-\hat{\bm{U}}\hat{\bm{\Psi}}\bm{\beta})^{\mathrm{H}}(\bm{r}-\hat{\bm{\Phi}}\bm{\alpha}-\hat{\bm{U}}\hat{\bm{\Psi}}\bm{\beta})}{q_\beta} \\ -\bm{\alpha}^{\mathrm{H}}\hat{\bm{\Gamma}}_\alpha\bm{\alpha}
\end{multline}
where $\propto^{\mathrm{e}}$ denotes proportionality after taking the exponential, and $\bm{\Gamma} = \mathrm{Diag}{(\bm{\gamma})}$. For the remaining expectation  we have, using the shorthand $\bm{r}_\beta = \bm{r}-\hat{\bm{\Phi}}\bm{\alpha}$,
\vspace*{-1mm}
\begin{align}
&\expect{(\bm{r}_\beta-\hat{\bm{U}}\hat{\bm{\Psi}}\bm{\beta})^{\mathrm{H}}(\bm{r}_\beta-\hat{\bm{U}}\hat{\bm{\Psi}}\bm{\beta})}{q_\beta} \nonumber\\ &= \bm{r}^{\mathrm{H}}_\beta \bm{r}_\beta - \bm{r}^{\mathrm{H}}_\beta \hat{\bm{U}}\hat{\bm{\Psi}}\hat{\bm{\beta}} - \hat{\bm{\beta}}^{\mathrm{H}}\hat{\bm{\Psi}}^{\mathrm{H}}\hat{\bm{U}}^{\mathrm{H}}\bm{r}_\beta + \expect{\bm{\beta}^{\mathrm{H}}\hat{\bm{\Psi}}^{\mathrm{H}}\hat{\bm{U}}^{\mathrm{H}}\hat{\bm{U}}\hat{\bm{\Psi}}\bm{\beta}}{q_\beta} \,. \\[-7mm]\nn
\end{align}
We can further rewrite this, as
\begin{align}
\expect{\bm{\beta}^{\mathrm{H}}\hat{\bm{\Psi}}^{\mathrm{H}}\hat{\bm{U}}^{\mathrm{H}}\hat{\bm{U}}\hat{\bm{\Psi}}\bm{\beta}}{q_\beta}&=\expect{(\bm{\beta}-\hat{\bm{\beta}})^{\mathrm{H}}\hat{\bm{\Psi}}^{\mathrm{H}}\hat{\bm{U}}^{\mathrm{H}}\hat{\bm{U}}\hat{\bm{\Psi}}(\beta-\hat{\bm{\beta}})}{q_\beta} \nonumber \\ 
&\hspace*{15mm}+ \hat{\bm{\beta}}^{\mathrm{H}}\hat{\bm{\Psi}}^{\mathrm{H}}\hat{\bm{U}}^{\mathrm{H}}\hat{\bm{\beta}} \\[-7mm]\nn
\end{align}
and
\begin{align}
& \expect{(\bm{\beta}-\hat{\bm{\beta}})^{\mathrm{H}} \hat{\bm{\Psi}}^{\mathrm{H}}\hat{\bm{U}}^{\mathrm{H}}\hat{\bm{U}}\hat{\bm{\Psi}}(\beta-\hat{\bm{\beta}})}{q_\beta} \nn\\ 
& \quad\quad = \mathrm{tr}{\left(\hat{\bm{U}}\hat{\bm{\Psi}} \expect{(\bm{\beta}-\hat{\bm{\beta}})(\bm{\beta}-\hat{\bm{\beta}})^{\mathrm{H}}}{q_\beta} \hat{\bm{\Psi}}^{\mathrm{H}}\hat{\bm{U}}^{\mathrm{H}}\right)} \nn\\ 
& \quad\quad = \mathrm{tr}{\left(\hat{\bm{U}}\hat{\bm{\Psi}} \hat{\bm{C}}_\beta \hat{\bm{\Psi}}^{\mathrm{H}}\hat{\bm{U}}^{\mathrm{H}}\right)} \,. \label{eq:expectation_trace} \\[-7mm]\nn
\end{align}
The trace term of~\eqref{eq:expectation_trace} is independent of $\bm{\alpha}$ and can be neglected, hence defining $\hat{\bm{r}}_\alpha = \bm{r}-\hat{\bm{U}}\hat{\bm{\Psi}}\hat{\bm{\beta}}$ and recasting the rest of the terms in the form of a complex normal log-\ac{pdf} yields
\vspace*{-1mm}
\begin{align}
\log q_\alpha &\propto^{\mathrm{e}} -\hat{\lambda}(\hat{\bm{r}}_\alpha-\hat{\bm{\Phi}\alpha})^{\mathrm{H}}(\hat{\bm{r}}_\alpha-\hat{\bm{\Phi}\alpha}) -\bm{\alpha}^{\mathrm{H}}\hat{\bm{\Gamma}}_\alpha\bm{\alpha} \nonumber\\ &= -\bm{\alpha}^{\mathrm{H}}\left( \hat{\lambda}\hat{\bm{\Phi}}^{\mathrm{H}}\hat{\bm{\Phi}} + \hat{\bm{\Gamma}}_\alpha \right) \bm{\alpha} +2\Re(\hat{\bm{r}}^{\mathrm{H}}_\alpha \hat{\bm{\Phi}} \bm{\alpha}) \nonumber\\ &{\hspace{0.5cm}} - \hat{\lambda}^2 \hat{\bm{r}}^{\mathrm{H}}_\alpha \hat{\bm{\Phi}} {\left( \hat{\lambda}\hat{\bm{\Phi}}^{\mathrm{H}}\hat{\bm{\Phi}} + \hat{\bm{\Gamma}}_\alpha \right)}^{-1} \hat{\bm{\Phi}}^{\mathrm{H}} \hat{\bm{r}}_\alpha \nonumber\\ &{\hspace{0.5cm}} + \hat{\lambda}^2 \hat{\bm{r}}^{\mathrm{H}}_\alpha \hat{\bm{\Phi}} {\left( \hat{\lambda}\hat{\bm{\Phi}}^{\mathrm{H}}\hat{\bm{\Phi}} + \hat{\bm{\Gamma}}_\alpha \right)}^{-1} \hat{\bm{\Phi}}^{\mathrm{H}} \hat{\bm{r}}_\alpha \nonumber\\ &= \log \text{CN}(\bm{\alpha} \vert \hat{\bm{\alpha}}, \hat{\bm{C}}_{\alpha})  + \log \vert \hat{\bm{C}}_{\alpha} \vert \nonumber\\ &{\hspace{0.5cm}} + \hat{\lambda}^2 \hat{\bm{r}}^{\mathrm{H}}_\alpha \hat{\bm{\Phi}} \hat{\bm{C}}_{\alpha} \hat{\bm{\Phi}}^{\mathrm{H}} \hat{\bm{r}}_\alpha
\label{eq:log_qalpha} \\[-7mm]\nn
\end{align}
where by coefficient comparison we obtain the result
\begin{align}
q(\bm{\alpha}) &= \text{CN}(\bm{\alpha} \vert \hat{\bm{\alpha}}, \hat{\bm{C}}_{\alpha}) \label{eq:update_alpha} \\
\hat{\bm{\alpha}} &= \hat{\lambda}\hat{\bm{C}}_{\alpha}\hat{\bm{\Phi}}^{\mathrm{H}}\hat{\bm{r}}_{\alpha} \label{eq:update_alpha_hat} \\
\hat{\bm{C}}_{\alpha} &= {\left( \hat{\lambda} \hat{\bm{\Phi}}^{\mathrm{H}}\hat{\bm{\Phi}} + \hat{\bm{\Gamma}}_{\alpha} \right)}^{-1} \,. \label{eq:update_C_alpha} \\[-7mm]\nn
\end{align}

The derivations for the interference channel weights $\bm{\beta}$ are analogous, leading to
\begin{align}
q(\bm{\beta}) &= \text{CN}(\bm{\beta} \vert \hat{\bm{\beta}}, \hat{\bm{C}}_{\beta}) \\
\hat{\bm{\beta}} &= \hat{\lambda}\hat{\bm{C}}_{\beta}\hat{\bm{\Psi}}^{\mathrm{H}}\hat{\bm{U}}^{\mathrm{H}}\hat{\bm{r}}_{\beta} \label{eq:update_beta_hat} \\
\hat{\bm{C}}_{\beta} &= {\left( \hat{\lambda} \hat{\bm{\Psi}}^{\mathrm{H}}\hat{\bm{U}}^{\mathrm{H}}\hat{\bm{U}}\hat{\bm{\Psi}} + \hat{\bm{\Gamma}}_{\beta} \right)}^{-1} \,. \label{eq:update_C_beta} \\[-7mm]\nn
\end{align}

\subsection{Update of Noise Precision}
Similarly for the noise precision update, noting only $\lambda$-dependent terms we have
\begin{equation}
\begin{split}
\log q_\lambda &\propto^{\mathrm{e}} (e_0+PN-1)\log\lambda-f_0\lambda \\ & -\lambda \expect{\expect{(\bm{r}-\hat{\bm{\Phi}}\bm{\alpha}-\hat{\bm{U}}\hat{\bm{\Psi}}\bm{\beta})^{\mathrm{H}}(\bm{r}-\hat{\bm{\Phi}}\bm{\alpha}-\hat{\bm{U}}\hat{\bm{\Psi}}\bm{\beta})}{q_\beta}}{q_\alpha} \,. \\[-7mm]\nn
\end{split}
\end{equation}
\vspace*{3mm}
The expectations are analogous to the one in~\eqref{eq:log_qalpha_1}, denoting $\hat{\bm{r}}_\lambda = \bm{r}-\hat{\bm{\Phi}}\hat{\bm{\alpha}}-\hat{\bm{U}}\hat{\bm{\Psi}}\hat{\bm{\beta}}$, leading to
\begin{equation}
\begin{split}
& \log q_\lambda \\ & \quad \propto^{\mathrm{e}} (e_0+PN-1)\log\lambda-f_0\lambda \\ & \quad\quad -\lambda \bigl( \hat{\bm{r}}_\lambda^{\mathrm{H}}\hat{\bm{r}}_\lambda + \mathrm{tr}{\bigl(\hat{\bm{\Phi}} \hat{\bm{C}}_\alpha \hat{\bm{\Phi}}^{\mathrm{H}}\bigr)} + \mathrm{tr}{\bigl(\hat{\bm{U}}\hat{\bm{\Psi}} \hat{\bm{C}}_\beta \hat{\bm{\Psi}}^{\mathrm{H}}\hat{\bm{U}}^{\mathrm{H}}\bigr)}  \bigr)
\end{split}
\end{equation}
and therefore
\begin{align}
{\hspace{-0.25cm}} q(\lambda) &= \text{Ga}(\lambda \vert e,f) \\
\hat{\lambda} = \frac{e}{f} &= \dfrac{e_{0}+PN}{f_{0}+ \hat{\bm{r}}^{\mathrm{H}}_\lambda \hat{\bm{r}}_\lambda + \mathrm{tr}(\hat{\bm{\Phi}}\hat{\bm{C}}_{\alpha}\hat{\bm{\Phi}}^{\mathrm{H}}) + \mathrm{tr}{(\hat{\bm{U}}\hat{\bm{\Psi}} \hat{\bm{C}}_\beta \hat{\bm{\Psi}}^{\mathrm{H}}\hat{\bm{U}}^{\mathrm{H}})}} \,. \label{eq:update_lambda_hat} \\[-7mm]\nn
\end{align}
For the hyper-parameters of the Gamma distribution we let $e_0 = f_0 = 0$ that is the non-informative Jeffery's improper hyper-prior.

\subsection{Estimations of Parameters}

The cost function for the object channel dispersion parameters $\bm{\zeta}$ is $Z_\alpha$ as in~\eqref{eq:Z_i}. For optimization, the equivalent $\log Z_\alpha$ will be used. The expectation term is identical to the solution for $q_\alpha$, except that $\bm{\Phi(\zeta)}$-dependent terms are to be explicitly considered. The result in~\eqref{eq:log_qalpha} contains all these non-constant terms. Then, marginalization of $\bm{\alpha}$ integrates out $\text{CN}(\bm{\alpha} \vert \hat{\bm{\alpha}}, \hat{\bm{C}}_{\alpha})$ to one as it is a valid \ac{pdf}. I.e., 
\begin{align}
\log Z_\alpha &\propto^{\mathrm{e}} \log \int \exp \Bigl( \log \text{CN}(\bm{\alpha} \vert \hat{\bm{\alpha}}, \hat{\bm{C}}_{\alpha})  + \log \vert \hat{\bm{C}}_{\alpha} \vert \nn\\[-1mm] &{\hspace{2.5cm}} + \hat{\lambda}^2 \hat{\bm{r}}^{\mathrm{H}}_\alpha \hat{\bm{\Phi}} \hat{\bm{C}}_{\alpha} \hat{\bm{\Phi}}^{\mathrm{H}} \hat{\bm{r}}_\alpha \Bigr) \mathrm{d}\bm{\alpha} \nn\\[1mm]
&= \log \vert \hat{\bm{C}}_{\alpha} \vert + \hat{\lambda}^2 \hat{\bm{r}}^{\mathrm{H}}_\alpha \hat{\bm{\Phi}} \hat{\bm{C}}_{\alpha} \hat{\bm{\Phi}}^{\mathrm{H}} \hat{\bm{r}}_\alpha \nn\\ &{\hspace{2.5cm}} + \log \int \text{CN}(\bm{\alpha} \vert \hat{\bm{\alpha}}, \hat{\bm{C}}_{\alpha}) \mathrm{d}\bm{\alpha}
\end{align}
and hence the result
\begin{equation}\label{eq:update_zeta}
\hat{\bm{\zeta}} = \argmax_{\bm{\zeta}} \Bigl( \log \vert \hat{\bm{C}}_{\alpha}(\bm{\zeta}) \vert + \hat{\lambda}^2 \hat{\bm{r}}^{\mathrm{H}}_\alpha \bm{\Phi}(\bm{\zeta}) \hat{\bm{C}}_{\alpha}(\bm{\zeta}) \bm{\Phi}^{\mathrm{H}}(\bm{\zeta}) \hat{\bm{r}}_\alpha \Bigr)
\end{equation}
where we explicitly noted all $\bm{\zeta}$-dependencies including inside the covariance matrix estimate $\hat{\bm{C}}_{\alpha}(\bm{\zeta})$.
This result can be recast to express the cost function for the parameters of a single component $\bm{\zeta}_l$ with all other parameters fixed. We order the components so that the index $l$ is last and denote all others by $\bar{l}$, i.e., $\bm{\Phi}(\bm{\zeta}) = \bm{\Phi}({[\bm{\zeta}_{\bar{l}}  \, \bm{\zeta}_{l}]}^{\mathrm{T}}) = [ \bm{\Phi}_{\bar{l}} \, \bm{\phi}_l ]$. Then, $\hat{\bm{C}}_{\alpha}$ can be written as the block matrix
\begin{align}
\hat{\bm{C}}_{\alpha} &= 
{\begin{bmatrix}
\hat{\bm{C}}_{\alpha,\bar{l}}^{-1} & \hat{\lambda}\bm{\Phi}_{\bar{l}}^{\mathrm{H}}\bm{\phi}_l \\
\hat{\lambda}\bm{\phi}_{l}^{\mathrm{H}}\bm{\Phi}_{\bar{l}} & \hat{c}_{\alpha,l}^{-1} \\
\end{bmatrix}}^{-1} \\
&= 
\begin{bmatrix}
\hat{\bm{C}}_{\alpha,\bar{l}} \,+ & -\hat{\lambda}\hat{\bm{C}}_{\alpha,\bar{l}}\bm{\Phi}_{\bar{l}}^{\mathrm{H}}\bm{\phi}_l\hat{c}_{\alpha,l}^{\prime} \\[-3pt]
{\hspace{3pt}}\hat{\lambda}^{2}\hat{\bm{C}}_{\alpha,\bar{l}}\bm{\Phi}_{\bar{l}}^{\mathrm{H}}\bm{\phi}_l\hat{c}_{\alpha,l}^{\prime}\bm{\phi}_{l}^{\mathrm{H}}\bm{\Phi}_{\bar{l}}\hat{\bm{C}}_{\alpha,\bar{l}} & {} \\[6pt]
-\hat{\lambda}\hat{c}_{\alpha,l}^{\prime}\bm{\phi}_{l}^{\mathrm{H}}\bm{\Phi}_{\bar{l}}\hat{\bm{C}}_{\alpha,\bar{l}} & \hat{c}_{\alpha,l}^{\prime} \\
\end{bmatrix} \notag
\end{align}
where we have applied the formula for block matrix inversion and defined $\hat{\bm{C}}_{\alpha,\bar{l}} = {(\hat{\lambda}\bm{\Phi}_{\bar{l}}^{\mathrm{H}}\bm{\Phi}_{\bar{l}} +  \hat{\bm{\Gamma}}_{\alpha,\bar{l}})}^{-1}$, $\hat{c}_{\alpha,l} = {(\hat{\lambda}\bm{\phi}_{l}^{\mathrm{H}}\bm{\phi}_{l} + \hat{\gamma}_{\alpha,l})}^{-1}$ and $\hat{c}_{\alpha,l}^{\prime} = {(\hat{c}_{\alpha,l}^{-1} - \hat{\lambda}^2\bm{\phi}_{l}^{\mathrm{H}}\bm{\Phi}_{\bar{l}}\hat{\bm{C}}_{\alpha,\bar{l}}\bm{\Phi}_{\bar{l}}^{\mathrm{H}}\bm{\phi}_l )}^{-1}$. \vspace*{0.5mm} In~\eqref{eq:update_zeta}, for the log-determinant of the block matrix we simply obtain $\log \vert \hat{\bm{C}}_{\alpha} \vert = \log \vert \hat{c}_{\alpha,l}^{\prime} \vert + \log \vert \hat{\bm{C}}_{\alpha,\bar{l}} \vert$. $\hat{\bm{C}}_{\alpha,\bar{l}}$ is independent of $\bm{\zeta}_l$ and hence can be neglected in the resulting cost function. Concerning the other term of~\eqref{eq:update_zeta}, noting that
\begin{align}
\hspace{-0.1cm} \bm{\Phi} \hat{\bm{C}}_{\alpha} \bm{\Phi}^{\mathrm{H}} &= \bm{\Phi}_{\bar{l}} \hat{\bm{C}}_{\alpha,\bar{l}} \bm{\Phi}_{\bar{l}}^{\mathrm{H}} + \hat{\lambda}^2 \hat{c}_{\alpha,l}^{\prime} \bm{\Phi}_{\bar{l}} \hat{\bm{C}}_{\alpha,\bar{l}} \bm{\Phi}_{\bar{l}}^{\mathrm{H}} \bm{\phi}_{l} \bm{\phi}_{l}^{\mathrm{H}} \bm{\Phi}_{\bar{l}} \hat{\bm{C}}_{\alpha,\bar{l}} \bm{\Phi}_{\bar{l}}^{\mathrm{H}} \nn\\
&{\hspace{0.1cm}} - 2 \Re \bigl( \hat{\lambda} \hat{c}_{\alpha,l}^{\prime} \bm{\Phi}_{\bar{l}} \hat{\bm{C}}_{\alpha,\bar{l}} \bm{\Phi}_{\bar{l}}^{\mathrm{H}} \bm{\phi}_{l} \bm{\phi}_{l}^{\mathrm{H}} \bigr)  + \bm{\phi}_{l} \hat{c}_{\alpha,l}^{\prime} \bm{\phi}_{l}^{\mathrm{H}} \nn\\[1mm]
&= \bm{\Phi}_{\bar{l}} \hat{\bm{C}}_{\alpha,\bar{l}} \bm{\Phi}_{\bar{l}}^{\mathrm{H}} \\
&{\hspace{0.1cm}} + {(\hat{\lambda} \bm{\Phi}_{\bar{l}} \hat{\bm{C}}_{\alpha,\bar{l}} \bm{\Phi}_{\bar{l}}^{\mathrm{H}} - \mathbf{I})}^{\mathrm{H}} \bm{\phi}_{l} \hat{c}_{\alpha,l}^{\prime} \bm{\phi}_{l}^{\mathrm{H}} {(\hat{\lambda} \bm{\Phi}_{\bar{l}} \hat{\bm{C}}_{\alpha,\bar{l}} \bm{\Phi}_{\bar{l}}^{\mathrm{H}} - \mathbf{I})} \nn
\end{align}
with the first term being a constant, we finally arrive at the result
\begin{multline}\label{eq:update_zeta_l}
\hat{\bm{\zeta}}_l = \argmax_{\bm{\zeta}_l} \Bigl( \log \vert \hat{c}_{\alpha,l}^{\prime}(\bm{\zeta}_l) \vert \\ + \hat{\lambda}^2 \hat{c}_{\alpha,l}^{\prime}(\bm{\zeta}_l) {\| \bm{\phi}_{l}^{\mathrm{H}}(\bm{\zeta}_l) {(\hat{\lambda} \bm{\Phi}_{\bar{l}} \hat{\bm{C}}_{\alpha,\bar{l}} \bm{\Phi}_{\bar{l}}^{\mathrm{H}} - \mathbf{I})} \hat{\bm{r}}_\alpha \|}^2 \Bigr) \,.
\end{multline}

For the interference chirp parameter estimates the derivations are analogous to~\eqref{eq:update_zeta} for every ramp independently, i.e.,
\begin{multline}\label{eq:update_theta_p}
\hat{\bm{\theta}} = \argmax_{\bm{\theta}} \Bigl( \log \vert \hat{\bm{C}}_{\beta}(\bm{\theta}) \vert \\ + \hat{\lambda}^2 \hat{\bm{r}}^{\mathrm{H}}_\beta \bm{U}(\bm{\theta}) \hat{\bm{\Psi}} \hat{\bm{C}}_{\beta}(\bm{\theta}) \hat{\bm{\Psi}}^{\mathrm{H}} \bm{U}(\bm{\theta})^{\mathrm{H}} \hat{\bm{r}}_\beta \Bigr)
\end{multline}
where the dependence of variables on the ramp index $p$ has been left out to reduce notational clutter.

\subsection{Fast Updates of Component Precisions}
As discussed in Section~\ref{sec:algorithm:fastupdate}, for our proposed algorithm we also apply the ``fast update rule'' for components as derived in~\cite{shutin2011TSP:fastVSBL}. Applying~\eqref{eq:mean_field_update} yields
\begin{align}
\log q_{\gamma_{\alpha,l}} &\propto^{\mathrm{e}} a_0 \log \gamma_{\alpha,l} - (b_0 + \expect{{\vert \alpha_l \vert}^2}{q_{\alpha}}) \gamma_{\alpha,l} \\
&\propto^{\mathrm{e}} a_0 \log \gamma_{\alpha,l} - (b_0 + {\vert \hat{\alpha}_l \vert}^2 + \hat{\bm{C}}_{\alpha}[l,l]) \gamma_{\alpha,l}
\end{align}
which leads to the solution
\begin{align}
q(\gamma_{\alpha,l}) &= \text{Ga}(\gamma_{\alpha,l} \vert a_l,b_l) \\
\hat{\gamma}_{\alpha,l} = \frac{a_l}{b_l} &= \frac{a_{0}+1}{b_{0}+\left( {\vert \hat{\alpha}_l \vert}^2 + \hat{\bm{C}}_{\alpha}[l,l] \right)} \label{eq:update_gamma_l_hat}
\end{align}
where we take $a_0 = b_0 = 0$. However, we can further investigate the behavior of this implicit equation when repeatedly updating $q_{\alpha}$ and $q_{\gamma_{\alpha,l}}$. We insert~\eqref{eq:update_alpha_hat} into~\eqref{eq:update_gamma_l_hat} and recast the resulting terms to ``isolate'' the $l$-th component as already shown for the derivation of~\eqref{eq:update_zeta_l}. I.e.,
\begin{align}
\hat{\gamma}_{\alpha,l}^{-1} &= {( \hat{\lambda}^2 \hat{\bm{C}}_\alpha \hat{\bm{\Phi}}^{\mathrm{H}} \hat{\bm{r}}_\alpha \hat{\bm{r}}_\alpha^{\mathrm{H}} \hat{\bm{\Phi}} \hat{\bm{C}}_\alpha^{\mathrm{H}} + \hat{\bm{C}}_\alpha )}[l,l] \notag \\
&= \hat{\lambda}^2 {\vert c_{\alpha,l}^{\prime} \hat{\bm{\phi}}_l^{\mathrm{H}} \hat{\bm{r}}_{\alpha} - \lambda c_{\alpha,l}^{\prime} \hat{\bm{\phi}}_l^{\mathrm{H}} \hat{\bm{\Phi}}_{\bar{l}} \hat{\bm{C}}_{\alpha,\bar{l}} \hat{\bm{\Phi}}_{\bar{l}}^{\mathrm{H}} \hat{\bm{r}}_{\alpha}  \vert}^2 + c_{\alpha,l}^{\prime}
\end{align}
and by defining~\eqref{eq:fast_rho} and~\eqref{eq:fast_omega}, and hence $c_{\alpha,l}^{\prime} = \hat{\gamma}_{\alpha,l}^{-1} \rho_l / (\hat{\gamma}_{\alpha,l}^{-1} + \rho_l)$, this can be further rewritten as
\begin{align}
\hat{\gamma}_{\alpha,l}^{-1} &= {\vert c_{\alpha,l}^{\prime} \vert}^2 \rho_l^{-2} \omega_l^2 + c_{\alpha,l}^{\prime} \notag \\
&= \dfrac{\hat{\gamma}_{\alpha,l}^{-2} (\omega_l^2 + \rho_l) + \hat{\gamma}_{\alpha,l}^{-1} \rho_l^2}{ {\vert \hat{\gamma}_{\alpha,l}^{-1} + \rho_l \vert}^2 }\,. \label{eq:fast_gamma_map}
\end{align}
By analyzing~\eqref{eq:fast_gamma_map} in the context of iterative updates as a rational map, it can be found that its stationary points are located at $\{0,\omega_l^2-\rho_l\}$. It can furthermore be proven that the non-zero stationary point is reached (i.e., the estimate $\hat{\gamma}_{\alpha,l}$ converges) if and only if $\rho_l/\omega_l^2 > T$ with $T=1$, which leads to the thresholding scheme with threshold $T$ (see Algorithm~\ref{alg:FastUpdate}). For a detailed discussion see \cite{buchgraber_variational_2013,shutin2011TSP:fastVSBL}. The component precisions of the interference channel $\hat{\gamma}_{\beta,k}\,$ can be found in a similar manner.

\section{Measurement of Interference}
\label{sec:appendix:meas_data}

As introduced in Section~\ref{subsec:meas_ex}, a measurement scheme whereby a separate interference signal is made available was devised. The parameters of the interference measurement can be found as Int. Measurement in Table~\ref{tab:params}. The measurement took place in an indoor office with the victim and interfering sensors on a desk opposite each other. Crucially, the scenario was therefore completely static with no movement of objects. First, the location and duration of the interference bursts received by the victim radar were identified manually. The static object signal could then be estimated by taking the sample mean of the non-interfered data. This estimate was subtracted from the measured signal, approximately removing the object signal, leaving interference and measurement noise. Then, the time-limited interference bursts were cut out to exclude irrelevant noise-only sections.

In Section~\ref{sec:signal-model}, the effect of the \ac{aaf} in the sensor hardware on the resulting interference burst was discussed (see the term $\bar{G}(f)$ in~\eqref{eq:IF_signal_interference}). This is typically assumed to be known to some precision in the application by the algorithm designers of an automotive radar manufacturer. Section~\ref{subsec:robustness_model_error} investigated the resilience of the proposed algorithm to a minor inaccuracy of this kind in the inference model. However, because the used hardware was a radar evaluation board from an external supplier, no details on the transfer function of the system was accessible to the authors. This can lead to a \textit{major} model error. To mitigate this, the signal was additionally filtered digitally with a filter whose magnitude response dominates that of the sensor hardware. The concatenation of a high- and a low-pass was applied, both being 8-th order Butterworth filters with respective cut-off frequencies of $0.5\,\mathrm{MHz}$ and $5\,\mathrm{MHz}$. The known response of this digital filter chain was then used in our inference model, noting that this still constitutes a rather significant model error. Finally, the interference signal prepared in this way was added onto the measurement from Obj. Measurement, yielding our test radar signal.

\bibliographystyle{IEEEtran}
\bibliography{IEEEabrv,references}

\vfill

\end{document}